\documentclass[acmsmall]{acmart}

\usepackage{hyperref}
\usepackage{graphicx, subcaption}
\usepackage{diagbox}
\usepackage{enumitem}
\usepackage{multirow}

\usepackage{xcolor}

\newlist{Properties}{enumerate}{2}
\setlist[Properties]{label=Property \arabic*., font=\textbf, itemindent=*}

\newcommand{\ie}{\emph{i.e., }}
\newcommand{\eg}{\emph{e.g., }}

\newcommand{\wrt}{\emph{w.r.t. }}
\newcommand{\cf}{\emph{cf. }}

\newcommand{\Lapl}{\mathbf{\mathop{\mathcal{L}}}}

\newcommand{\EE}{\mathop{\mathbb{E}}}
\newcommand{\VV}{\mathop{\mathbb{V}}}

\newcommand{\Mat}[1]{\boldsymbol{#1}}
\newcommand{\Set}[1]{\mathcal{#1}}

\AtBeginDocument{%
  }

\begin{document}

\setcopyright{acmlicensed}
\acmJournal{TOIS}
\acmYear{2023} \acmVolume{1} \acmNumber{1} \acmArticle{1} \acmMonth{1}
% \acmPrice{15.00}
\acmDOI{10.1145/3637061}

%%
%% The "title" command has an optional parameter,
%% allowing the author to define a "short title" to be used in page headers.
\title{On the Effectiveness of Sampled Softmax Loss for Item Recommendation}

%%
%% The "author" command and its associated commands are used to define
%% the authors and their affiliations.
%% Of note is the shared affiliation of the first two authors, and the
%% "authornote" and "authornotemark" commands
%% used to denote shared contribution to the research.
% \authornote{Both authors contributed equally to this research.}
\author{Jiancan Wu}
\email{wujcan@gmail.com}
\orcid{0000-0002-6941-5218}
\affiliation{%
  \institution{MoE Key Laboratory of Brain-inspired Intelligent Perception and Cognition, University of Science and Technology of China}
  \streetaddress{443 Huangshan Road}
  \city{Hefei}
  \country{China}
  \postcode{230027}
}

\author{Xiang Wang}
\authornotemark[2]
\authornote{Xiang Wang is also affiliated with Institute of Artificial Intelligence, Institute of Dataspace, Hefei Comprehensive National Science Center.}
\email{xiangwang1223@gmail.com}
\orcid{0000-0002-6148-6329}
\affiliation{%
  \institution{MoE Key Laboratory of Brain-inspired Intelligent Perception and Cognition, University of Science and Technology of China}
  \streetaddress{443 Huangshan Road}
  \city{Hefei}
  \country{China}
  \postcode{230027}
}

\author{Xingyu Gao}
\authornote{Xiang Wang and Xingyu Gao are corresponding authors.}
% \authornotemark[1]
\email{gxy9910@gmail.com}
\orcid{0000-0002-4660-8092}
\affiliation{%
  \institution{Institute of Microelectronics, Chinese Academy of Sciences}
  \city{Beijing}
  \country{China}
  \postcode{100029}
}

\author{Jiawei Chen}
\email{sleepyhunt@zju.edu.cn}
\orcid{0000-0002-4752-2629}
\affiliation{%
  \institution{Zhejiang University}
  \city{Hangzhou}
  \country{China}
  \postcode{310058}
}

\author{Hongcheng Fu}
\email{darrenfu@tencent.com}
\orcid{0009-0000-0956-0811}
\affiliation{%
  \institution{Tencent Music Entertainment Group}
  \city{Shenzhen}
  \country{China}
  \postcode{518000}
}

\author{Tianyu Qiu}
\email{timmyqiu@tencent.com}
\orcid{0009-0002-7095-8831}
\affiliation{%
  \institution{Tencent Music Entertainment Group}
  \city{Shenzhen}
  \country{China}
  \postcode{518000}
}

%%
%% By default, the full list of authors will be used in the page
%% headers. Often, this list is too long, and will overlap
%% other information printed in the page headers. This command allows
%% the author to define a more concise list
%% of authors' names for this purpose.
\renewcommand{\shortauthors}{Wu et al.}

%%
%% The abstract is a short summary of the work to be presented in the
%% article.
\begin{abstract}
The learning objective plays a fundamental role to build a recommender system.
% By casting the item recommendation problem as a supervised learning task, 
Most methods routinely adopt either pointwise (\eg binary cross-entropy) or pairwise (\eg BPR) loss to train the model parameters,
% while rarely pay attention to softmax loss due to the high computational cost.
while rarely pay attention to softmax loss, which assumes the probabilities of all classes sum up to 1, due to its computational complexity when scaling up to large datasets or intractability for streaming data where the complete item space is not always available.
The sampled softmax (SSM) loss emerges as an efficient substitute for softmax loss.
Its special case, InfoNCE loss, has been widely used in self-supervised learning and exhibited remarkable performance for contrastive learning.
Nonetheless, limited recommendation work uses the SSM loss as the learning objective.
Worse still, none of them explores its properties thoroughly and answers ``Does SSM loss suit for item recommendation?'' and ``What are the conceptual advantages of SSM loss, as compared with the prevalent losses?'', to the best of our knowledge.

In this work, we aim to offer a better understanding of SSM for item recommendation.
Specifically, we first theoretically reveal three model-agnostic advantages:
(1) mitigating popularity bias, which is beneficial to long-tail recommendation;
(2) mining hard negative samples, which offers informative gradients to optimize model parameters;
and (3) maximizing the ranking metric, which facilitates top-$K$ performance.
However, based on our empirical studies, we recognize that the default choice of cosine similarity function in SSM limits its ability in learning the magnitudes of representation vectors. As such, the combinations of SSM with the models that also fall short in adjusting magnitudes (\eg matrix factorization) may result in poor representations.
One step further, we provide mathematical proof that message passing schemes in graph convolution networks can adjust representation magnitude according to node degree, which naturally compensates for the shortcoming of SSM.
Extensive experiments on four benchmark datasets justify our analyses, demonstrating the superiority of SSM for item recommendation.
Our implementations are available in both TensorFlow\footnote{\url{https://github.com/wujcan/SSM-TensorFlow}} and PyTorch\footnote{\url{https://github.com/wujcan/SSM-Torch}}.
\end{abstract}

%%
%% The code below is generated by the tool at http://dl.acm.org/ccs.cfm.
%% Please copy and paste the code instead of the example below.
%%


%%
%% Keywords. The author(s) should pick words that accurately describe
%% the work being presented. Separate the keywords with commas.

% \received{20 February 2007}
% \received[revised]{12 March 2009}
% \received[accepted]{27 November 2023}

%%
%% This command processes the author and affiliation and title
%% information and builds the first part of the formatted document.
\maketitle

% !TeX root = ./0_main.tex
\section{Introduction}
% TODO: 1. learner remains largely unexplored. pointwise, pairwise, softmax loss;
In recent years, studies on recommendation modeling have been extensively conducted.
Many model architectures have been proposed to capture user preference from user-item interactions, covering multilayer perceptrons \cite{He2017NCF}, attention mechanisms \cite{NAIS}, generative models \cite{Yang23DreamRec}, and graph neural networks \cite{He2020LightGCN,PinSage,KipfW17GCN,WuYQS0023GIF}, etc.
However, relatively few studies focus on the learning objective --- the way to train and optimize the model parameters.
Specifically, most of work casts the item recommendation problem into a supervised learning task, and adopts one of the following learning objectives: 
% \begin{itemize}[leftmargin=*]
\begin{itemize}
    \item \textbf{Pointwise loss}. From the perspective of binary classification or regression, it treats the observed user-item interactions as positive instances, and other missing data as weak negatives.
    For a user-item pair, it encourages the model prediction to fit the corresponding label, so as to approach user preference directly.
    The common choices include binary cross-entropy \cite{He2017NCF,Shan2016Deep,WuHWWCLX22GCM} and mean square error \cite{Koren2008Factorization,He2017NFM}.
    
    \item \textbf{Pairwise loss}. It models the relative ordering of a positive item over an unobserved item for a target user.
    % Formally, for a user, it enforces the recommender to score the observed item higher than the unobserved.
    Numerically, it forces the model to score an observed item higher than its unobserved counterparts. 
    Representative pairwise losses are BPR \cite{Rendle2009BPR,Wang2019Neural,Wan00WGT22CPR} and WARP~\cite{Weston2011WSABIE}.
    % Among them, BPR is one of the most popular and effective losses in recommendation.
    
    \item \textbf{Softmax loss}. By applying a softmax function, it normalizes the model predictions over all items as a probability distribution. It then maximizes the probability of the observed items as compared with that of the unobserved items.
    Although aligned well with the ranking metrics that emphasize the top ranked positions \cite{Rendle2021Item,Bruch2019An}, softmax loss is relatively little used in recommender systems.
    % One reason is in its time complexity --- in practice, the scale of items easily reaches millions or even larger, making the training cost unaffordable.
    One reason is that softmax loss assumes the probabilities of all items for each user sum up to 1, which may be too computationally expensive to scale to large datasets --- in practice, the scale of items easily reaches millions or even larger. Additionally, calculating the exact probability of each possible interaction may be intractable in streaming environments~\cite{Yi2019Sampling,GuoYWCZH19Streaming} where the complete item space is not always available.
\end{itemize}

\textbf{Sampled softmax (SSM)} emerges as a substitute for softmax.
% The basic idea is sampling a small set of negatives and employing the partition function of softmax loss on this subset, instead of all items.
The basic idea is to use a sampled subset of negatives instead of all items.
As such, it not only inherits the desired property of ranking, but also reduces the training cost dramatically~\cite{Covington2016Deep}.
However, directly optimizing classical recommendation models such as MF~\cite{Koren2009Matrix} may not yield satisfactory performance (\cf Table~\ref{Table:overall_comparion}).
Typically, existing recommendation work leverages SSM mainly for two purposes:
(1) Approximating the softmax function. Prior work \cite{Bengio2003Quick} argues that SSM loss is a biased version of full softmax loss. One possible solution is the log$Q$ correction \cite{Bengio2003Quick}, which samples negative instances from the softmax distribution.
Some follow-on efforts \cite{Wang2021Cross,Rawat2019Sampled,Yang2020Mixed,Yi2019Sampling,Bai2017TAPAS,Blanc2018Adaptive,Lian2020Personalized} devise different methods to reduce the sampling bias.
(2) Performing contrastive learning. Inspired by its success in CV \cite{Chen2020A,Gidaris2018Unsupervised,Oord2018Representation} and NLP \cite{Gao2021SimCSE,Devlin2019BERT,Lan2020ALBERT} domains, researchers are exploring contrastive learning for recommendation \cite{Zhou2021Contrastive,Wu2021Self,Zhou2020S3-Rec}.
They typically use the InfoNCE loss \cite{Oord2018Representation} for the auxiliary task,
% which maximizes the agreement between different views of the same users or items as compared with that
maximizing the agreement of positive pairs as compared with that of negative pairs.
% Interestingly, when viewing the observed and unobserved user-item pairs as positive and negative pairs respectively, InfoNCE loss becomes sampled softmax loss.
Essentially, InfoNCE is an SSM function, since the observed and unobserved user-item pairs can be viewed as positive and negative instances, respectively.

Nonetheless, only very limited studies \cite{Covington2016Deep,Zhou2021Contrastive,yao2021selfsupervised} utilize SSM as the main learning objective for model training.
Even though some recommendation work has used it, they do not explore its properties, failing to answer the questions ``Does SSM suit item recommendation?'' and ``What are the conceptual advantages of SSM, as compared with pointwise and pairwise losses?''.

In this work, we seek to better understand SSM loss for item recommendation.
% Through theoretical analyses, we observe
Firstly, we conduct theoretical analyses, identifying
three \textbf{model-agnostic advantages} of SSM loss:
% \begin{itemize}[leftmargin=*]
\begin{itemize}
    \item \textbf{Alleviating popularity bias.}
    % \textcolor{red}{Its approximated closed-form solution \wrt interactions is nearly inversely proportional to item popularity in log-scale, when we sample negative samples based on their frequency.}
    Equipped with in-batch negative sampling, its approximated closed-form solution \wrt interactions is inversely proportional to item popularity in log scale, naturally suppressing the prediction scores of popular items.
    
    \item \textbf{Mining hard negative samples.} It can discover hard negative samples, which contribute more informative and larger gradients to the optimization and representation learning.
    
    \item \textbf{Maximizing the ranking metric.} Optimizing SSM loss is consistent with maximizing the Discounted Cumulative Gain (DCG) metric, which is more suitable for the top-$K$ task.
\end{itemize}

We further conduct experiments to justify the superiority of SSM for item recommendation. Specifically, among collaborative filtering (CF) family, we select four classical models for studying: matrix factorization (MF) \cite{Koren2009Matrix} representing ID-based methods, SVD++ \cite{Koren2008Factorization} being the representation of history-based approaches, while NGCF~\cite{Wang2019Neural} and LightGCN~\cite{He2020LightGCN} on behalf of graph-based models, respectively.
We have two observations:
(1) In both normal and long-tail testing scenarios, history-based and graph-based models achieve leading performance, validating the usefulness of SSM; 
(2) To our surprise, when applying SSM on MF, we observe large performance degradation in most cases.
% Empirically, we use SSM loss to train a high-performing recommender, LightGCN \cite{He2020LightGCN}, on four benchmark datasets.
% Experimental results validate the superiority of SSM loss over the prevalent losses, in both normal and long-tail test scenarios.
% We further investigate how SSM loss behaves when applied on different types of models, so as to uncover its \textbf{model-specific characteristics}.
% Without loss of generality, among collaborative filtering (CF) family, we select matrix factorization (MF) \cite{Koren2009Matrix}, SVD++ \cite{Koren2008Factorization}, and LightGCN to represent ID-, history-, and graph-based CF methods, respectively.
% Experiments show that the loss achieves significant performance improvements on SVD++ and LightGCN, but causes performance drop for MF.
Scrutinizing these models, we find that MF has no mechanism to adapt the representation magnitude, as compared with SVD++, NGCF, and LightGCN.
Hence, we hypothesize that SSM may be good at learning the \textbf{representation directions}, but falls short in adjusting the \textbf{representation magnitudes}.
Moreover, we give theoretical analyses of this hypothesis.
In contrast, the history-based and graph-based models can compensate for the weakness of SSM.

In a nutshell, our contributions are summarized as follows:
% \begin{itemize}[leftmargin=*]
\begin{itemize}
    \item We present a deep understanding of SSM and theoretically analyze its model-agnostic advantages: reducing popularity bias, mining hard negatives, and maximizing DCG metric.
    
    \item We point out the weakness of SSM loss in learning representation magnitudes, and show that history- and graph- based recommenders compensate for this weakness.
    
    \item Extensive experiments on four benchmark datasets justify the rationality of our analyses, demonstrating the superiority of SSM for item recommendation.
\end{itemize}
% !TeX root = ./0_main.tex
\section{SSM for Recommendation}
Suppose we have a dataset $\mathcal{D}$ that consists of
$\mathfrak{M}$ users, $\mathfrak{N}$ items, and $|\mathcal{D}|$ observed interactions.
The goal of item recommendation is to recommend a user with a list of items she may be interested in.
Noise contrastive estimation (NCE)~\cite{Gutmann2012Noise} treats this as a binary classification task, discriminating between the observed data and some artificially generated noises.
Specifically, for each observed interaction $(u,i)$, where item $i$ is in user $u$'s set of adopted items $\mathcal{P}_u$, we independently sample $N$ negative items $\mathcal{N} = \{ j_1, j_2, \cdots, j_N \}$ from a noise distribution $p_n$.
Let $\mathcal{I} = \{i\} \cup \mathcal{N}$ be the union of the positive item and the sampled negatives. We assign a binary label $C_k$ to each tuple $(u,k) \in {u} \times \mathcal{I}$, where $C_k=1$ if $k=i$ and 0 otherwise.
% Denoting $\mathcal{P}_u = \{i_1, i_2, \cdots, i_{|\mathcal{P}_u|}\}$ as the set of items user $u$ has adopted, for each observed interaction $(u,i), i\in \mathcal{P}_u$, we sample another set of items $\mathcal{N} = \{ j_1, j_2, \cdots, j_N \}$ from the pre-defined distribution $p_n$ as negative items.
% Denote by 
% % $\mathcal{I}=\{i,j_1,j_2,\cdots,j_N\}$ 
% $\mathcal{I} = \{i\} \cup \mathcal{N} =\{i,j_1,j_2,\cdots,j_N\}$,
% the union of two sets $\{i\}$ and $\mathcal{N}$, and assign to each tuple $(u,k)\in \{u\} \times \mathcal{I}$ a binary class label $C_k: C_k=1$ if $k=i$ and $C_k=0$ if $k \in \mathcal{N}$.
% Since the ideal probability density function is unknown, we model the class-conditional probability $p(\cdot|C_k=1)$ with $p_m(\cdot)$, that is,
% \begin{align}
%     p(u,k|C_k=1)&=p_m(u,k),\nonumber\\ p(u,k|C_k=0)&=p_n(k).
% \end{align}
% Since the true data distribution is unknown, we approximate the positive conditional distribution $p(u,k \vert C_k=1)$ with a model $p_m(u,k)$.
% The negative conditional distribution $p(u,k \vert C_k=0)$ remains the noise distribution $p_n(k)$.
The priors are $p(C_k=1) = 1/(1+N)$ and $p(C_k=0) = N/(1+N)$. Using Bayes' rule, the posterior probabilities are:
\begin{gather}
p(C_k=1 \vert u,k) = \frac{p(u,k \vert C_k=1)}{p(u,k \vert C_k=1) + N p(u,k \vert C_k=0)}, \\
p(C_k=0 \vert u,k) = \frac{N p(u,k \vert C_k=0)}{p(u,k \vert C_k=1) + N p(u,k \vert C_k=0)}.
\end{gather}
%
% The prior probabilities are $p(C_k=1)=1/(1+N)$ and $p(C_k=0)=N/(1+N)$. Therefore, the posterior probabilities for each class are:
% \begin{gather}
%     p(C_k=1|u,k)=\frac{p_m(u,k)}{p_m(u,k) + N p_n(k)},\\
%     p(C_k=0|u,k)=\frac{N p_n(k)}{p_m(u,k) + N p_n(k)}.
% \end{gather}
Without loss of generality, let $p(C_k=1|u,k)=\sigma(f(u,k))$, where $\sigma(\cdot)$ is the \textit{sigmoid} function, and $f(u,k)$ measures the affinity score between $u$ and $k$.
As such, we can obtain:
\begin{gather}\label{eq:positive_conditional_probability}
    p(u,k \vert C_k=1) = N p(u,k \vert C_k=0) \exp{(f(u,k))}.
\end{gather}
Note that among the $(N+1)$ tuples, only one is labeled 1. Our goal is to maximize the following joint conditional probability:
\begin{align}
    P &\overset{(\romannumeral1)}{=} \frac{p(u,i,j_1,\cdots,j_N|C_i=1,C_{j_1}=0,\cdots,C_{j_N}=0)}{\sum_{j\in \mathcal{I}} p(u,i,j_1,\cdots,j_N|C_i=0,C_{j_1}=0,\cdots,C_{j}=1,\cdots,C_{j_N}=0)}\nonumber\\
    &\overset{(\romannumeral2)}{=} \frac{p(u,i|C_i=1) \prod_{k\in\mathcal{I} \setminus \{i\}} p(u,k|C_k=0)}{\sum_{j \in \mathcal{I}} p(u,j|C_j=1) \prod_{l \in \mathcal{I} \setminus \{j\}} p(u,l|C_l=0)}\nonumber\\
    &\overset{(\romannumeral3)}{=}\frac{N p(u,i \vert C_i=0) \exp{(f(u,i))} \prod_{k\in\mathcal{I} \setminus \{i\}} p(u,k \vert C_k=0) }{\sum_{j \in \mathcal{I}} N p(u,j \vert C_j=0) \exp{(f(u,j))} \prod_{l \in \mathcal{I} \setminus \{j\}} p(u,l \vert C_l=0)} \nonumber\\
    % &=\frac{p_m(u,i)/p_n(i)}{\sum_{j \in \mathcal{I}} p_m(u,j)/p_n(j)} \nonumber\\ 
    &\overset{(\romannumeral4)}{=} \frac{\exp{(f(u,i))}}{\sum_{j \in \mathcal{I}}\exp{(f(u,j))}},
\end{align}
% In the third line, we divide a term $\prod_{k\in\mathcal{I}} p_n(k)$ on both numerator and denominator simultaneously. 
where $(\romannumeral1)$ measures the normalized conditional probability that the observed item $i$ is correctly labeled as 1 while the rest are labeled as 0 (cf. the numerator) against any other situation where some $j \in \mathcal{I}$ is labeled as 1 while the rest are labeled as 0 (cf. the denominator), $(\romannumeral2)$ follows from the independence assumption of the $(N+1)$ tuples, $(\romannumeral3)$ is the direct result of applying Equation~\eqref{eq:positive_conditional_probability}, and $(\romannumeral4)$ utilizes the fact that both the numerator and denominator share a common factor of $N \prod_{k \in \mathcal{I}} p(u,k | C_k=0)$.
Usually, the negative-logarithm is applied for numerical stability, resulting in the sampled-softmax loss function:
\begin{gather}
    \Lapl_{SSM} = -\frac{1}{|\mathcal{D}|} \sum_{(u,i)\in \mathcal{D}} \log \frac{\exp{(f(u,i))}}{\exp{(f(u,i))} + \sum\limits_{j \in \mathcal{N}}\exp{(f(u,j))}}.
\end{gather}
% \begin{align}
%     &\Lapl_{SSM} \nonumber\\ 
%     &= -\frac{1}{|\mathcal{D}|} \sum_{(u,i)\in \mathcal{D}} \log \left\{ \frac{\exp{(f(u,i))}}{\exp{(f(u,i))} + \sum\limits_{j \in \mathcal{N}}\exp{(f(u,j))}} \right\} + \lambda \left \| \Theta \right \|_2^2
% \end{align}
% where $\lambda$ controls the $L_2$ regularization strength, $\Theta$ is model parameter.
Here, we omit the regularization term for simplicity.
Typically, we use the mini-batch stochastic gradient descent method, \eg Adam~\cite{Kingma2014Adam}, to optimize model parameters.
To make full use of parallel computing of modern hardware, in-batch negative sampling~\cite{Hidasi2015Session} is commonly adopted, that is, treating positive items of other users in the same batch as the negatives.
In expectation, this is equivalent to sampling based on the empirical frequency of items in the dataset.
Common choices of the affinity function $f(u,i)$ are cosine similarity or inner product of the representations of $u$ and $i$ with a temperature coefficient.
Recently, \cite{Wu2021Self,Khosla2020Supervised} have demonstrated that cosine similarity with temperature coefficient is a better choice, since it endows SSM with the ability to mine hard negatives in a bounded interval. We will go deep into this property later.
We should emphasize that throughout this paper, unless otherwise stated, SSM is equipped with temperature-aware cosine similarity function and in-batch negative sampling.

\section{Theoretical Analyses}
\label{sec:theorecical_analyses}

% \section{Model-agnostic Advantages}
% \label{sec:model_agnostic_advantages}
% \wjc{In this section, we first prove in theory the three advantages of SSM loss for item recommendation.
% Then, we point out the potential shortcoming of SSM loss in representation learning and provide theoretical supporting evidence that message passing scheme can intrinsically adjust representation magnitude according to node degree, which compensates for the shortcoming of SSM loss in adjusting the magnitude of representations.}
In this section, we first conduct thorough theoretical analyses to reveal three advantages of SSM for item recommendation. We then identify the potential limitation of SSM in learning  representation magnitude, and show that message passing scheme can compensate for this limitation.
% Then, we conduct empirical experiments to verify its effectiveness on both normal and long-tail recommendation.

\subsection{Properties of SSM}
\subsubsection{\textbf{Alleviating Popularity Bias}}
\label{sssec:analysis_popularity_bias}
Intuitively, items with a larger frequency are more likely to be involved in a batch.
In other words, this is equivalent to choosing negative items from an empirical frequency sampler, in expectation.
As a consequence, popular items are prone to be penalized as negatives, preventing the model from recommending them to some degree~\cite{Yi2019Sampling}.
Here, we prove this statement theoretically.

Let $\Lapl_u$ be the loss on user $u$
and $l(u,i)$ be the individual loss term of positive sample $(u,i)$, then we have,
\begin{align}
    \Lapl_u 
    = \sum_{i \in \mathcal{P}_u} l(u,i) \nonumber 
    = \sum_{i \in \mathcal{P}_u} \left\{ - \log \frac{\exp{(f(u,i))}}{\exp{(f(u,i))} + \sum_{j\in\mathcal{N}}\exp{(f(u,j))}} \right\}.
\end{align}
The gradient of $\Lapl_u$ \wrt $f(u,i)$ is as follows:
\begin{align}
    &\frac{\partial }{\partial f(u,i)} \left\{ \sum_{k\in \mathcal{P}_u} \left[\log \left( 1 + \sum_{j\in\mathcal{N}}\exp{\left(f(u,j) - f(u,k)\right)} \right)\right] \right\} \nonumber \\
    &= \sum_{k\in \mathcal{P}_u} \frac{\partial }{\partial f(u,i)} \log \left[ 1 + N \EE_{j\sim p_n}\exp{(f(u,j)-f(u,k))} \right].
    % =& \frac{1 + N f_n(i)}{\exp{(f(u,i))} + N \mathop{\mathbb{E}}_{j\in p_n} \exp{(f(u,j))}} \exp{(f(u,i))} \nonumber \\ 
    % & + \sum_{k \in \mathcal{P}_u \setminus \{i\}} \frac{N f_n(k)}{\exp{(f(u,k))} + N \mathop{\mathbb{E}}_{j\in p_n} \exp{(f(u,j))}}\exp{(f(u,i))} - 1 \nonumber\\
    % \approx & \frac{1 + N f_n(i)}{ N \mathop{\mathbb{E}}\limits_{j\in p_n} \exp{(f(u,j))}} \exp{(f(u,i))} \nonumber \\
    % &+ \sum_{k \in \mathcal{P}_u \setminus \{i\}} \frac{N f_n(i)}{ N \mathop{\mathbb{E}}\limits_{j\in p_n} \exp{(f(u,j))}} \exp{(f(u,i))} - 1 \nonumber\\
    % =& \frac{1 + |\mathcal{P}_u| N f_n(i)}{N \mathop{\mathbb{E}}\limits_{j\in p_n} \exp{(f(u,j))}} \exp{(f(u,i))} - 1
\end{align}
Here, we use the law of large numbers, changing the sum operation to the expectation.
$p_n$ is the distribution of negative samples which is predefined.
Note that there are two cases where the term $f(u,i)$ is involved:
(1) $k=i$, that is, positive sample is $(u,i)$. Note that with probability $p_n(i)$, item $i$ is sampled as a negative item in this case. As such, the gradient of this part is
\begin{align}
\label{eq:pos_part}
    -1 + \frac{1 + N p_n(i) }{\exp{(f(u,i))} + N \EE_{j\sim p_n}\exp{(f(u,j))}} \exp{(f(u,i))}.
\end{align}
(2) $k \neq i$ but $j = i$, that is, only negative sample is $(u,i)$. In this case, the gradient is
\begin{align}
\label{eq:neg_part}
    \sum_{k \in \mathcal{P}_u \setminus \{i\}} \frac{ N p_n(i) \exp{(f(u,i))} }{\exp{(f(u,k))} + N \EE_{j\sim p_n}\exp{(f(u,j))}}.
\end{align}
By adding \eqref{eq:pos_part} and \eqref{eq:neg_part} together, we obtain the total gradient of $\Lapl_u$ \wrt $f(u,i)$. We further enforce the total gradient to be zero and obtain the nearly closed-form solution of $f(u,i)$:
% In the third equation, we use the fact that when $N \to \infty$,
% $\exp{(f(u,k))}$ $\ll N \EE_{j\in p_n}\exp{(f(u,j))},k\in\mathcal{P}_u$.
% Therefore, we obtain the nearly close-form solution
\begin{align}
    f^*(u,i)=\log \frac{N \mathop{\mathbb{E}}\limits_{j\sim p_n} \exp{(f(u,j))}}{ 1 + N |\mathcal{P}_u| p_n(i)}.
\end{align}
Here we use the fact that 
$\exp{(f(u,k))}$ $\ll N \EE_{j\sim p_n}\exp{(f(u,j))}, k\in\mathcal{P}_u$ when $N \to \infty$. Notice that the larger $p_n(i)$ is (\ie the more popular the item is), the smaller $f^{*}(u,i)$ will be. 
% This admits that SSM loss has the potential to suppress popularity bias.
This is in line with Inverse Propensity Weighting (IPW) methods that adjust the data distribution to be even by reweighting the training instances for bias reduction~\cite{SchnabelSSCJ16Recommendations,Chen2020Bias}.
Prior work~\cite{Zhou2021Contrastive} revealed that InfoNCE~\cite{Oord2018Representation} (which has a similar formula as SSM) and IPW share the same global optima when setting the proposal distribution to be the propensity score. However, to our knowledge, we are the first to derive the closed-form solution of SSM, providing more intuitive explanations for why SSM can alleviate popularity bias and further enabling flexible control over item popularity.
% Moreover, unlike the IPW methods that require an extra model to estimate the propensity score, SSM is free from
We will conduct quantitative experiments to verify the strength of SSM in suppressing popularity bias in Section~\ref{ssec:exp_long_tail}.

It is worth mentioning that besides the in-batch negative sampling strategy, one can adopt any other negative sampler to pursue more flexibility. For example, a naive sampler that samples $N$ negatives for each observed interaction independently from any predefined $p_n$, at the expense of training speed.
We will conduct experiments to study the impact of $p_n$ and $N$ in Sections~\ref{sssec:exp_negative_distribution} and~\ref{sssec:exp_number_negatives}, respectively.
% In this work, to understand the effect of distribution $p_n$ and the number of negative samples, we devise a simple variant of in-batch sampling strategy which introduces a customized negative sampler as the naive sampler does, but shares the sampled $N$ negatived samples across the positive pairs in the current batch as the in-batch negative sampling strategy does. For ease of implementation, we let $p_n(i)=f_i^{\beta}$ which adjusts between uniform sampler and frequency sampler, controlled by coefficient $\beta \in [0,1]$.
% Experimental details can be seen in Section~\ref{sssec:exp_negative_distribution} and~\ref{sssec:exp_number_negatives}.
% Other choices of $p_n$ are left as future work.

\subsubsection{\textbf{Hard Negative Mining}}
\label{sssec:analysis_hard_negative_mining}
In~\cite{Wu2021Self,Khosla2020Supervised}, the authors analyzed that the InfoNCE loss in self-supervised learning is able to perform hard negative mining.
Due to the similar formulae of SSM and InfoNCE, we find that SSM exhibits similar power in mining hard negative items when learning user representation.

Formally, we denote the final representations of user $u$ and item $i$ as $\Mat{z}_u$ and $\Mat{z}_i$ respectively and adopt cosine similarity with temperature coefficient $\tau$ to measure the agreement between $u$ and $i$, that is, $f(u,i)=\frac{\Mat{s}_u^\top \Mat{s}_i}{\tau}$, where $\Mat{s}_u$ and $\Mat{s}_i$ are the normalized representations, \ie $\Mat{s}_u = \frac{\Mat{z}_u}{\left \| \Mat{z}_u \right \|},\Mat{s}_i = \frac{\Mat{z}_i}{\left \| \Mat{z}_i \right \|}$.
Then the gradient of $l(u,i)$ \wrt the user representation $\Mat{z}_u$ is as follows~\footnote{We adopt the numerator layout notation here.}:

\begin{align}
    \frac{\partial l(u,i)}{\partial \Mat{z}_u}
    % = \frac{1}{\tau \left \| z_u \right \|}\Big\{c(i) + \sum_{j\in\mathcal{N}} c(j)\Big\},
    = \frac{c(i) + \sum_{j\in\mathcal{N}} c(j)}{\tau \left \| \Mat{z}_u \right \|},
\end{align}
where,
\begin{align}
    c(k) =&
    \begin{cases} 
        \left( \Mat{s}_k - (\Mat{s}_u^\top \Mat{s}_k) \Mat{s}_u \right)^\top \left( P_{uk} - 1 \right),  & \mbox{if } k=i \\
        \left( \Mat{s}_k - (\Mat{s}_u^\top \Mat{s}_k) \Mat{s}_u \right)^\top P_{uk}, & \mbox{if } k=j
    \end{cases}\\
    % = \left( s_i - (s_u^T s_i) s_u \right)^T \left( P_{ui} - 1 \right), \nonumber \\
    % c(j) &= \left( s_j - (s_u^T s_j) s_u \right)^T P_{uj},  \nonumber \\
    &P_{uk} = \frac{\exp(\Mat{s}_u^\top \Mat{s}_k/\tau)}{\sum_{l\in \mathcal{I}} \exp(\Mat{s}_u^\top \Mat{s}_l / \tau)}.
\end{align}
Then, we focus on the magnitude contribution of negative item $j\in\mathcal{N}$, \ie $\left \| c(j) \right\|$, which is proportional to the following term:
\begin{gather}\label{eq:negative-contribution-prop}
    g(x) = \sqrt{1 - x^2} \exp \left(\frac{x}{\tau}\right), x\in[-1,1],
\end{gather}
where $x=\Mat{s}_u^\top \Mat{s}_j$ denotes the cosine similarity between $\Mat{z}_u$ and $\Mat{z}_j$.
By scrutinizing $g(x)$, we have the following insights:
(1) when setting a proper $\tau$ (empirically in the range of 0.1 to 0.2), hard negative items which have larger $x$ offer larger gradients to guide the optimization, thus making learned representations more discriminative and accelerating the training process.
(2) when $x$ becomes large enough (\ie $x>\left(\sqrt{\tau ^2 + 4} -\tau \right) / 2$), $g(x)$ decreases sharply to 0 as $x \to 1$, which indicates that SSM has a nice property that weakens the impact of too-hard negatives since they may be potential positives.
It is worth noting that the key difference between our analysis in this work and previous studies~\cite{Wu2021Self} is that we specifically focus on the hard negative mining property within user-item interactions. Previous work has mainly centered around self-supervised learning techniques to find hard samples from augmented nodes of the same type (i.e., user or item nodes). In contrast, by exploring the potential of using hard negative mining within the context of user-item interactions, we aim to provide more direct insights into the relationship between users and items, which can improve our understanding of the underlying mechanisms of recommender systems.
Compared with other hard negative mining techniques like choosing hard negatives from top recommendation results of the current training state~\cite{Zhang2013Optimizing} or setting a margin to block the gradient of easy negative items~\cite{Mao2021SimpleX}, SSM uses a more graceful way through in-batch negative sharing strategy and control factor $\tau$, which scarcely increases training complexity.

\subsubsection{\textbf{Maximizing DCG}}
\label{sssec:analysis_maximizing_DCG}
Discounted Cumulative Grain (DCG) is a widely-adopted ranking metric which uses a graded relevance scale to calculate the utility score.
The contribution to the utility from a relevant item reduced logarithmically proportional to the position of the ranked list, which mimics the behavior of a user who is less likely to examine items at larger ranking position.
Formally, DCG is defined as follows:
\begin{align}
    DCG(\pi_{f_u},y) = \sum_{i=1}^{|\mathcal{I}|} \frac{2^{y_i}-1}{\log_2(1+\pi_{f_u}(i))},
\end{align}
where $\pi_{f_u}$ is a ranked list over $\mathcal{I}$ induced by $f$ for user $u$; $y$ is a binary indicator: $y_i=1$ if item $i$ has been interacted by $u$, otherwise $y_i=0$; $\pi_{f_u}(i)$ is the rank of $i$.
It is worth noting that, in practice, there may be more than one item that has been interacted by $u$ in $\mathcal{I}$.
Inspired by~\cite{Bruch2019An}, we derive the relationship between SSM loss and DCG:
\begin{align}
    \pi_{f_u}(i)
    &= 1 + \sum_{j\in \mathcal{I} \setminus \{i\}} \mathbb{I}(f(u,j) - f(u,i) > 0) \nonumber\\
    &\leq 1 + \sum_{j\in \mathcal{I} \setminus \{i\}} \exp(f(u,j) - f(u,i)) \nonumber\\
    &= \sum_{j\in \mathcal{I}} \exp(f(u,j) - f(u,i)).
\end{align}
Based on the fact $\mathbb{I}(x > 0) \leq \exp{(x)}$, we have:
% Here, we use the fact that $\mathbb{I}(x > 0) \leq \exp{(x)}$.
% Finally, we have
\begin{align}
    -\log DCG(\pi_{f_u},y) 
    &\leq -\log \frac{1}{\log \left( 1+\pi_{f_u}(i) \right)} \nonumber\\
    &\leq -\log \frac{1}{\pi_{f_u}(i)} \nonumber \\
    &\leq -\log \frac{\exp (f(u,i))}{\sum_{j\in \mathcal{I}} \exp(f(u,j))}.
\end{align}
As the fact $\log (1+x) \leq x$, we can safely get that:
% Here, we use the fact that $\log (1+x) \leq x$.
% Therefore,
$-\log DCG$ is a lower bound of SSM, and minimizing SSM is equivalent to maximizing DCG of $\mathcal{I}$.
Such finding provides insight that SSM matches well with item recommendation in terms of optimization goal, thus being suitable for optimizing the recommendation models.

\subsection{Potential Limitations of SSM}
\label{ssec:disadvantage_SSM}
% In this section, we first conduct experiments of SSM loss paired with different recommender models on different benchmark datasets.
% Based on the experimental results, we then study the performance discrepancies of SSM loss \wrt different recommenders.

\subsubsection{\textbf{SSM Fails in Learning Representation Magnitude}}
\label{sssec:analysis_SSM_magnitude}
In the above section, we have proven that SSM has three desirable properties for item recommendation. Consequently, a spontaneous question is: ``Can SSM lead to empirical performance gain across different recommender models?'' Towards answering this question, we select three categories of CF methods --- ID-, history-, and graph- based methods --- as backbone models to justify its superiority. The detailed results are presented in Section~\ref{sssec:exp_different_recommender}.
To our surprise, in spite of significant performance gains on both history-based (\ie SVD++~\cite{Koren2008Factorization}) and graph-based models (\ie NGCF~\cite{Wang2019Neural} and LightGCN~\cite{He2020LightGCN}), we observe large performance degradation on ID-based model (\ie MF~\cite{Rendle2009BPR}).
Such inconsistency drives us to probe the failure of SSM.
Scrutinizing these recommender models, we find that the key lies in adapting representation magnitudes during representation learning.
For SSM, due to the adoption of cosine similarity, the representation magnitudes will not affect the optimization process (we ignore the regularization term here), thus SSM will not guide the representation magnitude learning.
What's worse, ID-based model (\ie MF) also has no explicit mechanism to adapt the representation magnitude, resulting in poor quality of representations.
On the contrary, as we will prove later, message-passing-based models such as LightGCN can intrinsically adjust the representation magnitude according to the node degree, thus compensating for the limitation of SSM.
Hence, we argue that SSM may be good at learning the \textbf{representation directions}, but falls short in adjusting the \textbf{representation magnitudes}.

\subsubsection{\textbf{Message Passing can Adjust the Representation Magnitude}}
\label{sssec:analysis_message_passing}
Without loss of generality, we define the message passage rule in user-item interaction graph as follows:
\begin{gather}
    \Mat{q}_i^{'} = \sum_{u\in \mathcal{P}_i} \frac{1}{|\mathcal{P}_i|^{\alpha_0} |\mathcal{P}_u|^{\alpha_1}} \Mat{p}_u,
\end{gather}
where $\alpha_0$ and $\alpha_1$ are the normalization factors. In this work, we only analyze from the item side, since the user side can be derived in the same way.

% Before starting our derivation step by step, since the observed data in item recommendation usually exhibit a power-law distribution~\cite{Milojevic2010PowerLaw,Clauset2009Power}, we first recap some basic knowledge about the Pareto distribution~\cite{pareto1964cours}.

% The Pareto distribution is a power-law probability distribution that is originally applied to describing the distribution of wealth in a society, fitting the trend that a large portion of wealth is held by a small fraction of the population.
% Suppose $X$ is a random variable with a Pareto (type 1) distribution, then the probability that $X$ is greater than $x$ is given by the survival function:
% \begin{align}
%     \bar{F}(x) = Pr(X \geq x) = 
%     \begin{cases}
%         1-\left( \frac{x_m}{x} \right )^ \alpha & x \geq x_m, \\ 
%         1 & x < x_m.
%     \end{cases} 
% \end{align}
% Formally, the probability density function of $X$ is
% \begin{align}
%     f(x) =
%     \begin{cases}
%         \frac{\alpha x_m^{\alpha}}{x^{\alpha + 1}} & x \geq x_m, \\ 
%         0 & x < x_m.
%     \end{cases} 
% \end{align}
% where $x_m$ is the minimum possible value of $X$, and $\alpha$ is a positive parameter which is known as the tail index.

Suppose the degree of user node $|\mathcal{P}_u|$ and item node $|\mathcal{P}_i|$ follow the Pareto distribution~\cite{pareto1964cours} with parameter $(\alpha, x_m=1)$, where $\alpha$ is the shape parameter, $x_m$ is the minimum possible value of node degree (we set it to 1). $\Mat{p}_u^{(0)}$ is initialized with some distribution $\mathcal{D}(\mu_0, \sigma_0^2)$, that is
\begin{gather}
    \EE(|\mathcal{P}_u|) = \EE(|\mathcal{P}_i|) = \frac{\alpha}{\alpha-1},\nonumber\\
    \VV(|\mathcal{P}_u|) = \VV(|\mathcal{P}_i|) = \frac{\alpha }{(\alpha - 1)^2 (\alpha - 2)} \nonumber\\
    \EE(\Mat{p}_u) = \EE(\Mat{q}_i) = \mu_0,\nonumber\\
    \VV(\Mat{p}_u) = \VV(\Mat{q}_i) = \sigma_0^2
\end{gather}
where $\EE(\cdot)$ and $\VV(\cdot)$ are the expectation and variance, respectively.
Therefore, for a given item $i$, the expectation of the square of its magnitude $\Mat{q}_i^{'}$ is
% \begin{align}
%     \EE((q_i^{'})^2)
%     &= \EE\left( \left(\sum_{u\in \mathcal{P}_i} \frac{1}{|\mathcal{P}_i|^{\alpha_0} |\mathcal{P}_u|^{\alpha_1}} p_u \right)^2 \right)\nonumber\\ 
%     &= \frac{1}{|\mathcal{P}_i|^{2\alpha_0}} \EE\left( \left( \sum_{u\in \mathcal{P}_i} \frac{1}{|\mathcal{P}_u|^{\alpha_1}} p_u \right)^2  \right) \nonumber\\ 
%     &= \frac{1}{|\mathcal{P}_i|^{2\alpha_0}} \left\{ \VV \left( \sum_{u\in \mathcal{P}_i} \frac{1}{|\mathcal{P}_u|^{\alpha_1}} p_u \right) + \left[ \EE\left( \sum_{u\in \mathcal{P}_i} \frac{1}{|\mathcal{P}_u|^{\alpha_1}} p_u \right) \right]^2 \right\}\nonumber\\ 
%     &= \frac{1}{|\mathcal{P}_i|^{2\alpha_0}} \left\{ \sum_{u\in \mathcal{P}_i} \VV \left( \frac{1}{|\mathcal{P}_u|^{\alpha_1}} p_u \right)  + \left[  \sum_{u\in \mathcal{P}_i} \EE\left( \frac{1}{|\mathcal{P}_u|^{\alpha_1}} p_u \right) \right]^2 \right\}
% \end{align}
% Here, we omit the following term on the right hand side of the last line since it's equal to 0 under the assumption that $u$ and $v$ are two independent samples,
% \begin{gather}
%     \sum_{u,v \in \mathcal{P}_i,\ u\neq v} \text{Cov} \left( \frac{1}{|\mathcal{P}_u|^{\alpha_1}} p_u, \frac{1}{|\mathcal{N}_v|^{\alpha_1}} p_v \right) = 0
% \end{gather}
% \begin{gather}
% \label{eq:qi_square}
%     \EE((q_i^{'})^2) =
%     \frac{1}{|\mathcal{P}_i|^{2\alpha_0}} \left\{ \sum_{u\in \mathcal{P}_i} \VV \left( \frac{1}{|\mathcal{P}_u|^{\alpha_1}} p_u \right)  + \left[  \sum_{u\in \mathcal{P}_i} \EE\left( \frac{1}{|\mathcal{P}_u|^{\alpha_1}} p_u \right) \right]^2 \right\}
% \end{gather}
\begin{gather}
\label{eq:qi_square}
    \EE((\Mat{q}_i^{'})^2) =
    \frac{1}{|\mathcal{P}_i|^{2\alpha_0}} \left\{ \sum_{u\in \mathcal{P}_i} V_u  + \left[  \sum_{u\in \mathcal{P}_i} E_u \right]^2 \right\},
\end{gather}
where,
\begin{gather}
    V_u = \VV \left( \frac{1}{|\mathcal{P}_u|^{\alpha_1}} \Mat{p}_u \right),\quad
    E_u = \EE\left( \frac{1}{|\mathcal{P}_u|^{\alpha_1}} \Mat{p}_u \right)
\end{gather}
% We leave the detailed derivation of Equation~\eqref{eq:qi_square} in Appendix~\ref{appendix:proof_equation} due to the limited space.
Since $|\mathcal{P}_u|$ and $\Mat{p}_u$ are independent variables, we have
% \begin{align}
%     \VV \left( \frac{1}{|\mathcal{P}_u|^{\alpha_1}} p_u \right)
%     &= \EE \left( \frac{1}{|\mathcal{P}_u|^{2\alpha_1}} \right) \EE \left( p_u^2 \right) - \left[ \EE \left( \frac{1}{|\mathcal{P}_u|^{\alpha_1}} \right) \right]^2 \left[ \EE(p_u) \right]^2 \nonumber\\ 
%     &= \frac{\alpha}{\alpha + 2\alpha_1}\left( \sigma_0^2 + \mu_0^2 \right) - \left( \frac{\alpha}{\alpha + \alpha_1} \right)^2 \mu_0^2 \\
%     \sum_{u\in \mathcal{P}_i} \EE\left( \frac{1}{|\mathcal{P}_u|^{\alpha_1}} p_u \right)
%     &= \sum_{u\in \mathcal{P}_i}\EE\left( \frac{1}{|\mathcal{P}_u|^{\alpha_1}} \right) \EE\left( p_u \right) = |\mathcal{P}_i| \frac{\alpha}{\alpha + \alpha_1} \mu_0.
% \end{align}
\begin{align}
    V_u
    &= \EE \left( \frac{1}{|\mathcal{P}_u|^{2\alpha_1}} \right) \EE \left( \Mat{p}_u^2 \right) - \left[ \EE \left( \frac{1}{|\mathcal{P}_u|^{\alpha_1}} \right) \right]^2 \left[ \EE(\Mat{p}_u) \right]^2 \nonumber\\ 
    &= \frac{\alpha}{\alpha + 2\alpha_1}\left( \sigma_0^2 + \mu_0^2 \right) - \left( \frac{\alpha}{\alpha + \alpha_1} \right)^2 \mu_0^2,
\end{align}
\begin{align}    
    \sum_{u\in \mathcal{P}_i} E_u
    &= \sum_{u\in \mathcal{P}_i}\EE\left( \frac{1}{|\mathcal{P}_u|^{\alpha_1}} \right) \EE\left( \Mat{p}_u \right)\nonumber\\
    &= |\mathcal{P}_i| \frac{\alpha}{\alpha + \alpha_1} \mu_0.
\end{align}
Finally, we can obtain
\begin{align}
    \EE\left((\Mat{q}_i^{'})^2\right) = a |\mathcal{P}_i|^{1-2\alpha_0} + b|\mathcal{P}_i|^{2-2\alpha_0}
    \label{eq:magnitude_vs_degree}
\end{align}
where,
\begin{gather}
    a = \frac{\alpha}{\alpha + 2\alpha_1}\left( \sigma_0^2 + \mu_0^2 \right) - \left( \frac{\alpha}{\alpha + \alpha_1} \right)^2 \mu_0^2 > 0 \nonumber\\
    b = \frac{\alpha^2 \mu_0^2}{(\alpha + \alpha_1)^{2} } > 0.
\end{gather}
Specially, when $\alpha_0=\alpha_1=0.5$, \ie the case of LightGCN, $\EE\left( (\Mat{q}_i^{'})^2 \right)$ monotonically linearly increases as $|\mathcal{P}_i|$ increases.
We conduct empirical experiments to study the impact of different message-passing strategies (\ie different $\alpha_0$ and $\alpha_1$) in section~\ref{sssec:exp_message_passing}.
While we find an effective way to compensate for the limitation of SSM loss in learning magnitude, the optimal solution is still an open question.
% We should emphasize that Equation~\eqref{eq:magnitude_vs_degree} also give supporting evidence about the degree bias problem in GCN-based methods~\cite{Tang2020Investigating,Zhao2021Bilateral}: without calibration, GCNs are biased towards high-degree nodes since they have larger magnitude in expectation.

% !TeX root = ./0_main.tex
\section{Experiments}
\label{sec:overall_experiments}
We conduct extensive experiments and answer the following research questions:
\begin{itemize}
    \item \textbf{RQ 1:} Does SSM suit well for item recommendation?
    
    \item \textbf{RQ 2:} How does SSM perform \wrt long-tail recommendation, as compared with the existing losses?
    
    \item \textbf{RQ 3:} How do different components affect SSM?
\end{itemize}

\subsection{Experimental Setup}
\subsubsection{\textbf{Compared Losses}}
To justify the superiority of SSM on item recommendation, we compare it with diverse losses:
% \begin{itemize}[leftmargin=*]
\begin{itemize}
    \item BCE Loss: A widely used pointwise loss that is formulated as:
        \begin{gather}
            \Lapl_{BCE} = -\sum\limits_{(u,i)\in \mathcal{D}} \log \sigma(f(u,i)) - \sum\limits_{(u,j)\in \mathcal{D}^{-}} \log \hat{\sigma}(f(u,j)),
        \end{gather}
        where $\hat{\sigma}(x) = 1 - \sigma(x)$, $\Set{D}^{-}$ is the sampled subset of unobserved interactions. 
    
    \item BPR Loss: The standard objective function in vanilla NGCF and LightGCN, which encourages the prediction of positive item to be larger than its negative counterpart. The formal formulation is defined as follows:
        \begin{gather}
            \Lapl_{BPR} = -\frac{1}{|\mathcal{D}|}\sum_{u=1}^{\mathfrak{M}} \sum_{i \in \mathcal{P}_u,\  j\not\in \mathcal{P}_u} \log{\sigma ( f(u,i) - f(u,j) )}.
        \end{gather}
    where $\mathfrak{M}$ is the total number of users in the dataset.
    
    \item SM Loss: It is short for softmax loss which maximizes the probability of the observed items normalized over all items by softmax function, that is:
        \begin{gather}
            \Lapl_{SM} = -\frac{1}{|\mathcal{D}|} \sum_{(u,i)\in \mathcal{D}} \log \left\{ \frac{\exp{(f(u,i))}}{\sum_{j=1}^\mathfrak{N} \exp{(f(u,j))}} \right\},
        \end{gather}
    where $\mathfrak{N}$ is the total number of items in the dataset.
    
    \item CCL Loss~\cite{Mao2021SimpleX}: It's a contrastive loss proposed recently by maximizing the cosine similarity of positive pairs while minimizing the similarity of negative pairs below a certain margin:
    % The formula is as follows:
    \begin{gather}
        \Lapl_{CCL} = -\frac{1}{|\mathcal{D}|} \sum\limits_{(u,i)\in \mathcal{D}} \left[1 - f(u,i) + \frac{w}{|\mathcal{N}|} \sum\limits_{j\in \mathcal{N}}(f(u,j)-m)_{+} \right]
    \end{gather}
    where $(\cdot)_{+}$ indicates $\max(0,\cdot)$, $\Set{N}$ is a set of randomly sampled negative samples, $m$ is the margin for similarity score of negative samples, $w$ is a hyper-parameter to balance the loss terms of positive samples and negative samples.
\end{itemize}

We implement SSM and all compared losses on different categories of CF models, ranging from ID-based (\ie MF~\cite{Rendle2009BPR}) that directly projects the single ID of a user/item into a latent embedding, to history-based (\ie SVD++~\cite{Koren2008Factorization}) that takes into consideration the user’s historical interactions for user representation learning, to graph-based methods (\ie NGCF~\cite{Wang2019Neural} and LightGCN~\cite{He2020LightGCN}) 
% which achieves the state-of-the-art performance with a light graph convolution for training efficiency and generalization.
that achieve state-of-the-art performance by performing graph convolutions on the user-item graph.

% Detailed comparison between SSM loss and competing losses is attached in Appendix~\ref{appendix:discussion_losses}.
\subsubsection{\textbf{Datasets and Evaluation Metrics}}
We conduct experiments on four benchmark datasets: Gowalla~\cite{Wang2019Neural,He2020LightGCN}, Yelp2018~\cite{Wang2019Neural,He2020LightGCN}, Amazon-Book~\cite{Wang2019Neural,He2020LightGCN} and Alibaba-iFashion~\cite{Wu2021Self}.
Following~\cite{Wang2019Neural,He2020LightGCN}, we use the same 10-core setting for the first three datasets. For Alibaba-iFashion, as processed in~\cite{Wu2021Self}, we randomly sample 300k users and collect all their interactions over the fashion outfits.
The statistics of all four datasets are summarized in Table~\ref{Table:Statistics of the experimented data}.
We follow the same strategy described in~\cite{Wang2019Neural} to split the interactions into training, validation and test set with a ratio of $7:1:2$. For users in the test set, we follow the all-ranking protocol~\cite{Wang2019Neural,He2020LightGCN} to evaluate the top-$K$ recommendation performance and report the average Recall@20 and NDCG@20.

\begin{table}[t]
\centering
\caption{Statistics of the datasets.}
% \vspace{-10pt}
\label{Table:Statistics of the experimented data}
\begin{tabular}{c|r|r|r|r}
\hline
Dataset          & \multicolumn{1}{c|}{\#Users} & \multicolumn{1}{c|}{\#Items} & \multicolumn{1}{c|}{\#Interactions} & \multicolumn{1}{c}{Density} \\ \hline \hline
Gowalla          & 29,858                       & 40,981                       & 1,027,370                           & 0.00084                     \\ \hline
Yelp2018         & 31,831                       & 40,841                       & 1,666,869                           & 0.00128                     \\ \hline
Amazon-Book      & 52,643                       & 91,599                       & 2,984,108                           & 0.00062                     \\ \hline
Alibaba-iFashion & 300,000                      & 81,614                       & 1,607,813                           & 0.00007                     \\ \hline
\end{tabular}
% \vspace{-10pt}
\end{table}

\subsubsection{\textbf{Hyper-parameter Settings}}
For fair comparison, all methods are trained from scratch which are initialized with Xavier~\cite{Glorot2010Understanding}.
In line with NGCF and LightGCN, we fix the embedding size to 64 and optimize all models via Adam~\cite{Kingma2014Adam} with the default learning rate of 0.001 and default mini-batch size of 2048.
The $L_2$ regularization term is added, with the coefficient in the range of $\{ 1e^{-6}, 1e^{-5}, \cdots, 1e^{-1} \}$. 
The normalization factor of SVD++ is searched in $\{0,0.5,1.0 \}$.
For graph-based method, \ie NGCG~\cite{Wang2019Neural} and LightGCN~\cite{He2020LightGCN}, the number of GCN layers $K$ is searched in the range of $\{ 1,2,3,4 \}$. The dropout ratio of NGCF is in $\{0.0, 0.1, · · · , 0.8\}$.
The layer combination coefficient of LightGCN is set to $\frac{1}{K+1}$.
For BCE, we randomly sample nonobserved interactions to form the negative set $\mathcal{D}^{-}$ in each training epoch, the ratio of positive to negative is set to $1:4$.
For BPR, we randomly sample a noninteracted item of the user as negative for each observed interaction.
For SM, we examine both cosine similarity and inner product similarity between user and item representations, and report the best performance.
% For CCL loss, since the authors have reported the results on Yelp2018 and Amazon-Book datasets under the same experimental setting, we just copy their results. While for the results of the rest two datasets which are not reported in the original paper~\cite{Mao2021SimpleX}, we report the best results based on our implementations.
For CCL, as suggested in the paper~\cite{Mao2021SimpleX}, we tune the margin $m$ among $0.1 \sim 1.0$ at an interval of 0.1 and weight $w$ in the range of $\{1, 150, 300, 1000 \}$. The number of negative samples is searched from 1 to 2048.
While for SSM, we find cosine similarity always leads to better performance.
Since the temperature coefficient $\tau$ in both SM loss and SSM is of great importance~\cite{Wu2021Self}, we use the following strategy to tune it:
we first perform grid search in a coarse grain range of $\{0.1, 0.2, 0.5, 1.0\}$, then in a finer grain range, which is based on the result of coarse tuning.
For example, if the model achieves the best performance when $\tau=0.2$ in the first stage, then we tune it in the range of $\{ \cdots, 0.16,0.18, 0.22, 0.24 \cdots \}$ at finer granularity.
The overall results reported in Table~\ref{Table:overall_comparion} are the average value of 5 runs.

\subsection{Overall Performance Comparison}
% We report the overall results in Table~\ref{Table:overall_comparion}.

\subsubsection{\textbf{Comparison Among Different Losses}}
\label{sssec:exp_different_losses}
Comparing the best performance each loss can achieve in Table~\ref{Table:overall_comparion},
we have the following observations:
% \begin{itemize}[leftmargin=*]
\begin{itemize}
    \item Among the compared losses, BCE performs worst on all four datasets, which suggests the limitations of pointwise objective.
    % \textcolor{red}{it usually assumes uses have concrete preference scores on items and builds a model to predict those scores while neglecting the diversity and variety of user preference.}
    Specifically, as it approaches the task as a binary classification machine learning task, fitting the exact value of labels may introduce noise in item recommendation.
    BPR performs better than BCE in most cases, verifying the superiority of pairwise objectives to capture relative relations among items.
    Both SM and CCL have significant gains over BCE and BPR on Yelp2018, Amazon-Book, and Alibaba-iFashion datasets.
    A common point of SM and CCL is that they both compare one positive sample with multiple negative samples.
    This suggests that enlarging the number of negative samples during training is beneficial to representation learning.
    On Gowalla, SM and CCL have on-par performance compared to BPR.
    % This shows the great potential of listwise objectives in mimicking the user's behaviors toward the recommended list.

    \item The best performance on each dataset is always achieved by SSM, empirically verifying the advantages of SSM for item recommendation.
    We attribute the gains to:
    (1) the use of multiple negatives for each observed interaction at every iteration, as compared with pointwise and pairwise loss.
    % Comparing with pointwise and pairwise loss, SSM uses multiple negatives for each observed interaction at every iteration.
    In addition, with a proper temperature coefficient, SSM benefits from dynamic hard negative mining (see Section~\ref{sssec:analysis_hard_negative_mining}), in which the hard negative items offer larger gradients to guide the optimization. Moreover, SSM has strong connections with DCG ranking metric as we have analyzed in Section~\ref{sssec:analysis_maximizing_DCG} --- meaning that SSM directly optimizes the goal of model evaluation.
    (2) using only a fraction of items from the whole item set for optimization, as compared with SM,
    % (2) Compared with SM, SSM loss only uses a fraction of items from the whole item set for optimization, 
    which greatly alleviates the training cost and difficulty, especially when the set of whole items becomes extremely large.
    Evidence supporting this assertion is that the improvements of SSM over SM become more significant on larger datasets (\ie Yelp2018, Amazon-Book, and Alibaba-iFahsion) than a smaller one (\ie Gowalla).
    Another possible reason lies in that by introducing randomness, SSM can avoid the phenomenon that a few extremely hard negative items dominate the optimization in SM.
    % Moreover, we conduct the significant test, where $p$-value < 0.05 indicates that the improvements of SSM loss over BPR loss are statistically significant in all cases.
    
    % \item It's worth mentioning that through our empirical experiments, we find that SSM loss converges much faster than BPR loss. We also ascribe such efficient training to the superiority of hard negative mining. On the contrary, for BPR loss, we randomly sample an unobserved item as negative, which is more likely to be an easy negative sample --- that is, the positive observation has larger predicted score than that of the selected negative sample and thus making the gradient magnitude vanishing~\cite{Rendle2014Improving}.
\end{itemize}

\begin{table*}[t]
\centering
\caption{Performance of different recommenders under different loss functions. The bold indicates the best result for each recommender on each dataset. The superscript $*$ indicates the best result on each dataset.}
% \vspace{-5pt}
\label{Table:overall_comparion}
\resizebox{0.99\textwidth}{!}{
\begin{tabular}{cc|cc|cc|cc|cc}
\hline
\multicolumn{2}{c|}{\textbf{Dataset}}                                   & \multicolumn{2}{c|}{\textbf{Gowalla}}         & \multicolumn{2}{c|}{\textbf{Yelp2018}}        & \multicolumn{2}{c|}{\textbf{Amazon-Book}}     & \multicolumn{2}{c}{\textbf{Alibaba-iFashion}} \\ \hline
\multicolumn{1}{c|}{\textbf{Recommender}}               & \textbf{Loss} & Recall                & NDCG                  & Recall                & NDCG                  & Recall                & NDCG                  & Recall                & NDCG                  \\ \hline
\multicolumn{1}{c|}{\multirow{5}{*}{\textbf{MF}}}       & \textbf{BCE}  & 0.1555                & 0.1294                & 0.0571                & 0.0464                & 0.0303                & 0.0236                & 0.0950                & 0.0439                \\
\multicolumn{1}{c|}{}                                   & \textbf{BPR}  & 0.1558                & 0.1254                & 0.0562                & 0.0454                & 0.0352                & 0.0266                & 0.1031                & 0.0483                \\
\multicolumn{1}{c|}{}                                   & \textbf{SM}   & 0.1777                & 0.1434                & \textbf{0.0709}       & \textbf{0.0581}       & 0.0515                & 0.0399                & 0.1225                & 0.0593                \\
\multicolumn{1}{c|}{}                                   & \textbf{CCL}  & \textbf{0.1837}       & \textbf{0.1493}       & 0.0698                & 0.0572                & \textbf{0.0559}       & \textbf{0.0447}       & \textbf{0.1229}       & \textbf{0.0585}       \\
\multicolumn{1}{c|}{}                                   & \textbf{SSM}  & 0.1231                & 0.0878                & 0.0509                & 0.0404                & 0.0473                & 0.0367                & 0.0841                & 0.0400                \\ \hline
\multicolumn{1}{c|}{\multirow{5}{*}{\textbf{SVD++}}}    & \textbf{BCE}  & 0.1549                & 0.1284                & 0.0564                & 0.0459                & 0.0306                & 0.0235                & 0.0969                & 0.0458                \\
\multicolumn{1}{c|}{}                                   & \textbf{BPR}  & 0.1589                & 0.1302                & 0.0569                & 0.0458                & 0.0373                & 0.0286                & 0.1094                & 0.0509                \\
\multicolumn{1}{c|}{}                                   & \textbf{SM}   & \textbf{0.1852}       & \textbf{0.1550}       & 0.0679                & 0.0555                & 0.0478                & 0.0371                & 0.1157                & 0.0559                \\
\multicolumn{1}{c|}{}                                   & \textbf{CCL}  & 0.1819                & 0.1453                & 0.0693                & 0.0567                & \textbf{0.0557}       & \textbf{0.0440}       & 0.1233                & 0.0591                \\
\multicolumn{1}{c|}{}                                   & \textbf{SSM}  & 0.1669                & 0.1376                & \textbf{0.0693}       & \textbf{0.0569}       & 0.0547                & 0.0429                & \textbf{0.1276}       & \textbf{0.0622}       \\ \hline
\multicolumn{1}{c|}{\multirow{5}{*}{\textbf{NGCF}}}     & \textbf{BCE}  & 0.1541                & 0.1303                & 0.0563                & 0.0463                & 0.0330                & 0.0264                & 0.0547                & 0.0238                \\
\multicolumn{1}{c|}{}                                   & \textbf{BPR}  & 0.1548                & 0.1248                & 0.0579                & 0.0477                & 0.0357                & 0.0273                & 0.0923                & 0.0416                \\
\multicolumn{1}{c|}{}                                   & \textbf{SM}   & 0.1766                & 0.1471                & 0.0699                & 0.0573                & 0.0486                & 0.0377                & 0.1169                & 0.0552                \\
\multicolumn{1}{c|}{}                                   & \textbf{CCL}  & 0.1778                & 0.1411                & 0.0651                & 0.0535                & 0.0516                & 0.0406                & 0.1068                & 0.0507                \\
\multicolumn{1}{c|}{}                                   & \textbf{SSM}  & \textbf{0.1854}       & \textbf{0.1548}       & \textbf{0.0736}       & \textbf{0.0608}       & \textbf{0.0552}       & \textbf{0.0431}       & \textbf{0.1296$^{*}$} & \textbf{0.0629$^{*}$} \\ \hline
\multicolumn{1}{c|}{\multirow{5}{*}{\textbf{LightGCN}}} & \textbf{BCE}  & 0.1743                & 0.1491                & 0.0628                & 0.0515                & 0.0373                & 0.0292                & 0.0983                & 0.0458                \\
\multicolumn{1}{c|}{}                                   & \textbf{BPR}  & 0.1824                & 0.1554                & 0.0640                & 0.0524                & 0.0417                & 0.0322                & 0.1086                & 0.0511                \\
\multicolumn{1}{c|}{}                                   & \textbf{SM}   & 0.1756                & 0.1429                & 0.0708                & 0.0580                & 0.0489                & 0.0377                & 0.1155                & 0.0558                \\
\multicolumn{1}{c|}{}                                   & \textbf{CCL}  & 0.1790                & 0.1407                & 0.0669                & 0.0554                & 0.0528                & 0.0416                & 0.1203                & 0.0570                \\
\multicolumn{1}{c|}{}                                   & \textbf{SSM}  & \textbf{0.1869$^{*}$} & \textbf{0.1571$^{*}$} & \textbf{0.0737$^{*}$} & \textbf{0.0609$^{*}$} & \textbf{0.0590$^{*}$} & \textbf{0.0459$^{*}$} & \textbf{0.1253}       & \textbf{0.0599}       \\ \hline
\end{tabular}
}
\end{table*}

\subsubsection{\textbf{Impact of Different Recommenders}}
\label{sssec:exp_different_recommender}
In the previous section, we have verified the superiority of SSM loss compared with other losses.
Here, we study if SSM loss is consistently a good choice for optimizing different CF models.
From Table~\ref{Table:overall_comparion}, we can find that:
% \begin{itemize}[leftmargin=*]
\begin{itemize}
    % \item Paired with SSM loss, both LightGCN and SVD++ have significant performance improvements over the combination of BCE and BPR loss. This again verifies the superiority of SSM loss for implicit item recommendation.
    % However, we observe performance drop when training MF via SSM loss.
    % as compared to BPR loss.
    % This admits that, SSM loss is more friendly to model history-based methods and graph-based methods, but may not suit for models like ID-based methods.
    % We argue that the key lies in message passing since both SVD++ and graph-based methods adopts message passing technique to refine representations.
    % We will make in-depth analyses on this problem later, from the perspective of learning representation magnitude.
    \item Paired with SSM, both history-based and graph-based methods achieve leading performance. This again verifies the superiority of SSM for item recommendation.
    However, when implemented on MF, SSM performs poorly. We attribute this phenomenon to the following two causes:
    (1) as analyzed in Section~\ref{ssec:disadvantage_SSM}, SSM neglects the impact of representation magnitudes when calculating matching score since it adopts cosine similarity function. Worse still, MF merely maps an ID into an embedding vector, without explicit design to adjust the magnitude of representations.
    As such, the combination of MF and SSM cannot compensate for the flaws in adjusting the magnitude of representations and hence leads to low-quality representation.
    % which is also incapable of adapting the magnitudes.
    % As such, the combination of MF and SSM loss may result in sub-optimal representation.
    (2) as analyzed in Section~\ref{sssec:analysis_popularity_bias}, SSM will penalize the predicted score for popular items. However, prior work~\cite{Zhang2021Causal} has demonstrated that popularity bias can be somehow good if properly leveraged. Given that MF has no mechanism to leverage popularity information, the combination of MF and SSM will under-estimate the popular items, leading to undesirable performance.
    In contrast, we have proven that message-passing methods inherently are capable of adjusting the representation magnitude based on node degree. Therefore, paired with a message-passing method, SSM will lead to excellent performance.
    
    % Based on the aforementioned empirical studies and a general idea that a high quality representation should have a proper direction and a proper magnitude,
    % we conjecture that SSM loss falls short in learning proper magnitude of representations, since it uses cosine similarity which cares only about the angle between two latent vectors while neglecting the impact of their magnitude.
    % Note that MF merely projects ID into latent embedding, it does not provide any design to adjust the magnitude of representations.
    % As such, the combination of MF and SSM loss can not compensate for the flaws in adjusting the magnitude of representations and hence performs poorly.
    % On the contrary, as we will prove in the following section, a message-passing method like SVD++ or GCN is capable of adjusting the magnitude of node representations inherently. Therefore, paired with a message-passing method, SSM loss will lead to excellent performance.
    
    \item Focusing on BPR or SSM only, we find that as the recommender becomes more complicated, the performance improves gradually. This is in line with the intuition that adding GCN layers will capture higher-order collaborative signal which is of benefit to recommendation.
    However, as we can see, the combination of SVD++ and SSM outperforms the combination of LightGCN and BPR. This suggests that BPR is less effective to mine user preference underlying observed interactions.
    In contrast, the performance gain provided by SSM could even be larger than that from adding GCN layers, which further justifies its effectiveness.
    
    \item Between the two graph-based recommenders, NGCF is less competitive than LightGCN when paired with traditional losses, \eg BCE and BPR. Prior studies typically attribute this to the adoption of multiple nonlinear transformations, which increases the difficulties to train the recommender well~\cite{He2020LightGCN, ChenWHZW20Revisiting}.
    However, armed with SSM loss, NGCF achieves an equivalent level (\eg on Gowalla and Yelp2018 datasets) or even better (\eg on Alibaba-iFashion dataset) performance compared with LightGCN. This indicates that SSM promotes the training process of more complicated models, showing great potential for solving complex optimization problems.
    
    %%
    % todo: discussion about different datasets?
    %%

\end{itemize}

\subsection{Long-tail Recommendation}
\label{ssec:exp_long_tail}
As analyzed in section~\ref{sssec:analysis_popularity_bias}, SSM is promising for alleviating the popularity bias. To verify this property, we follow~\cite{Wu2021Self}, splitting items into ten groups based on the frequency while keeping the total number of interactions of each group the same. The items in groups with larger GroupIDs have larger degrees.
As such, we decompose the Recall@20 metric of the whole dataset into ten parts, each of which represents the contribution of a single group as follows:

\begin{gather}\label{eq:group-recall}
    {\rm Recall} 
    = \frac{1}{\mathfrak{M}}\sum_{u=1}^{\mathfrak{M}}\frac{\sum\limits_{g=1}^{10} \left | (l_{rec}^u)^{(g)} \cap l_{test}^u \right |}{\left | l_{test}^u \right |} = \sum\limits_{g=1}^{10} {\rm Recall}^{(g)}
\end{gather}

where ${\rm Recall}^{(g)}$ measures the recommendation performance over the $g$-th group, $l_{rec}^{u}$ and $l_{test}^{u}$ are the items in the top-$K$ recommendation list and relevant items for user $u$, respectively.
We report the results in Fig.~\ref{fig:long-tail} and have the following observations:

\begin{figure*}[t]
 \centering
 \subcaptionbox{Gowalla\label{fig:long-tail-gowalla}}{
  \includegraphics[width=0.45\textwidth]{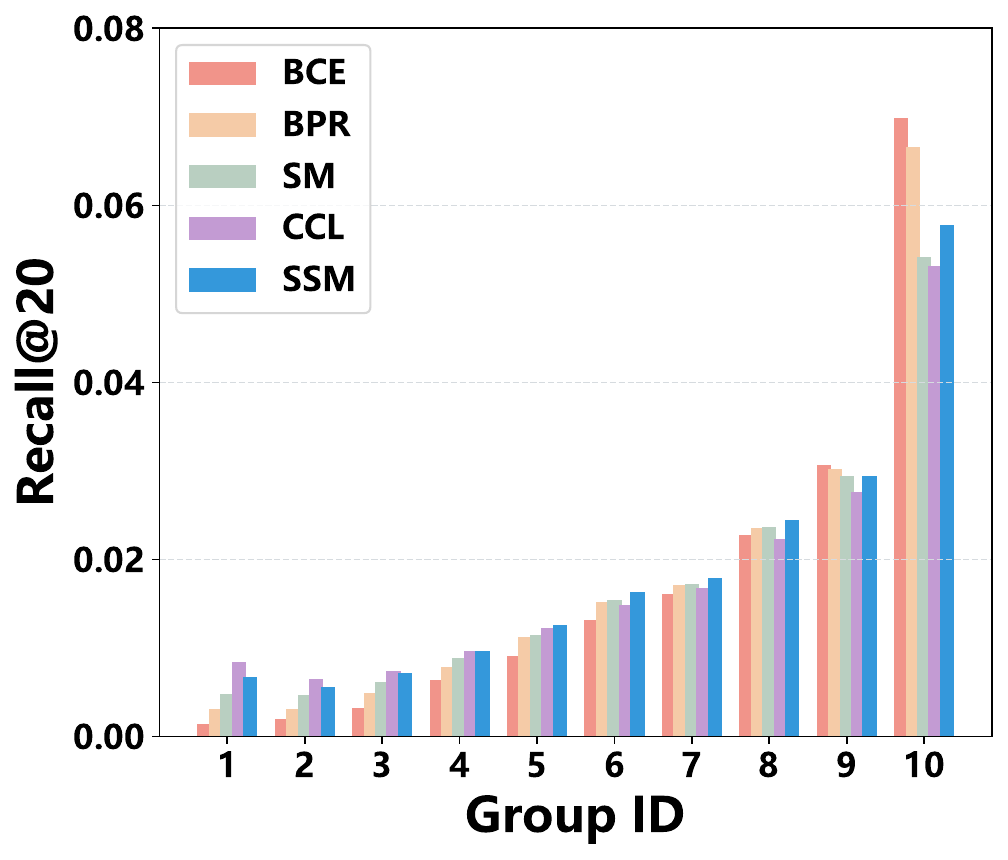}}
 \subcaptionbox{Yelp2018\label{fig:long-tail-yelp}}{
  \includegraphics[width=0.45\textwidth]{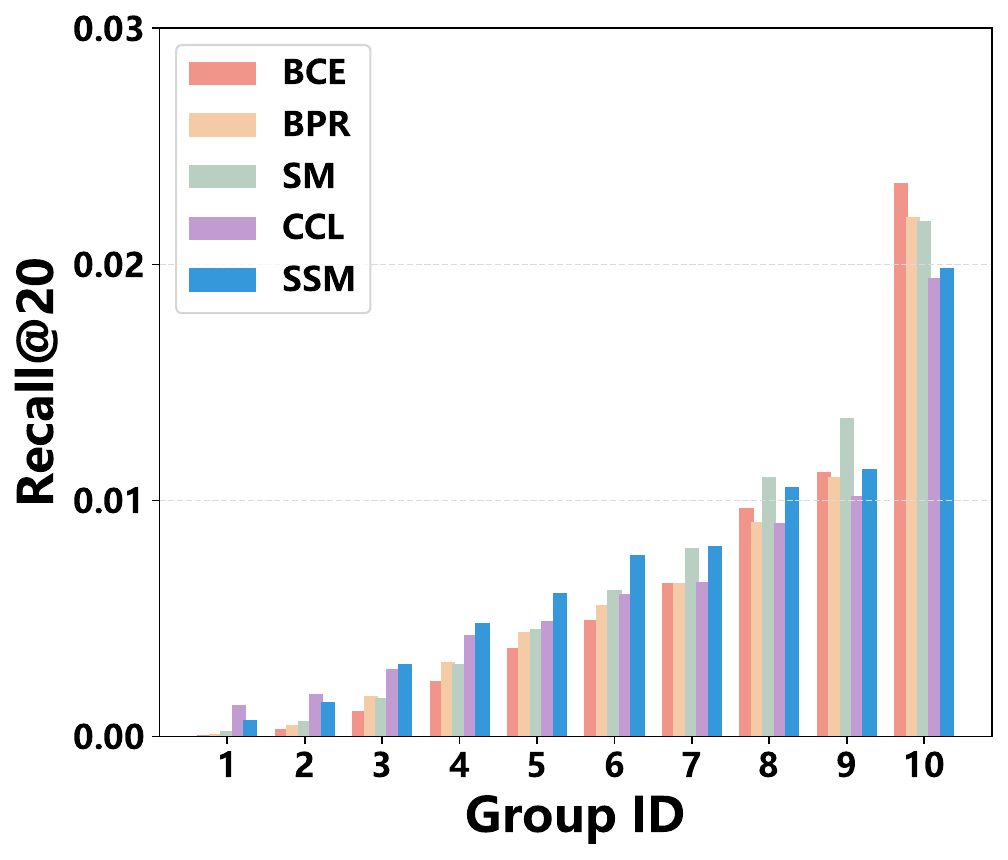}}
 \subcaptionbox{Amazon-Book\label{fig:long-tail-ab}}{
  \includegraphics[width=0.45\textwidth]{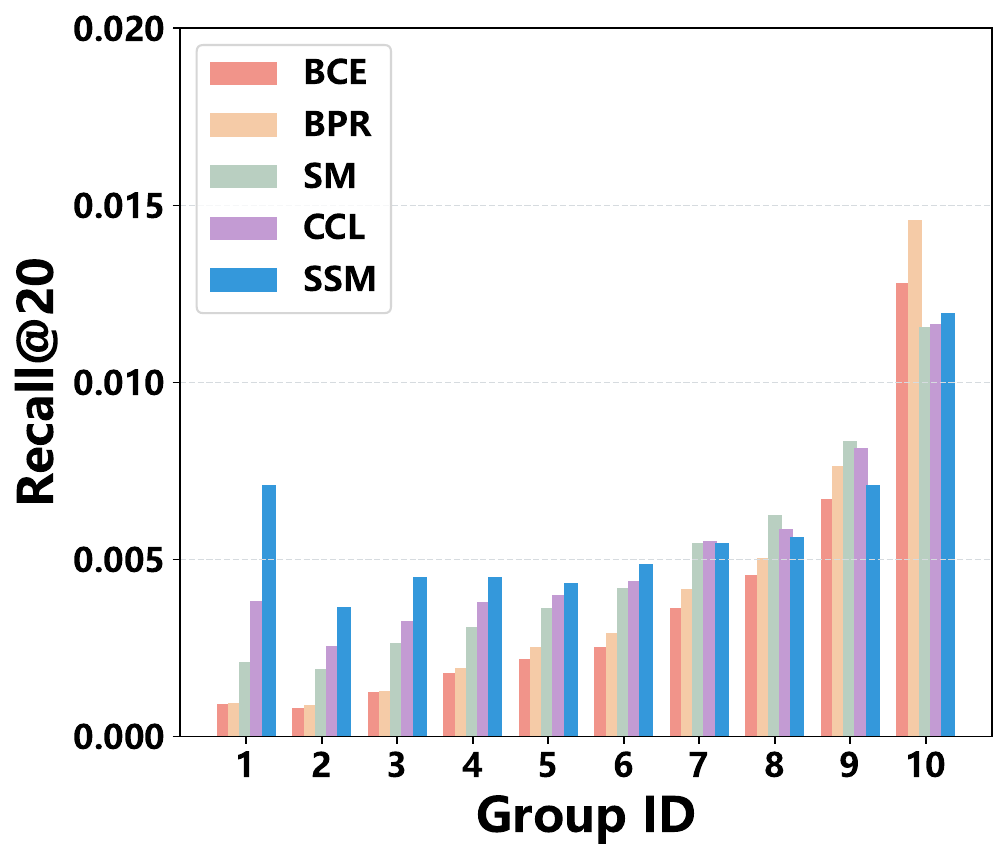}}
  \subcaptionbox{Alibaba-iFashion\label{fig:long-tail-ifashion}}{
  \includegraphics[width=0.45\textwidth]{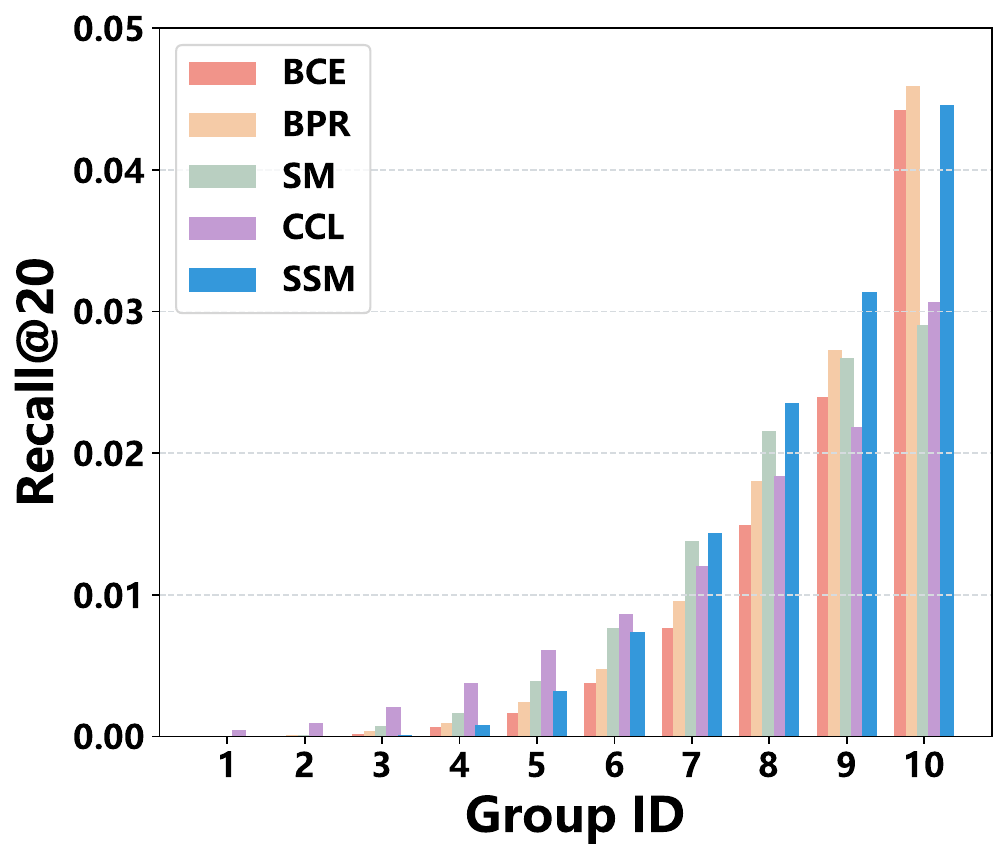}}
  % \vspace{-10pt}
  \caption{Performance comparison over different item groups among different objectives.}
 \label{fig:long-tail}
 % \vspace{-10pt}
\end{figure*}

\begin{itemize}
    \item Implemented on LightGCN, BCE, BPR, and SM all show a strong inclination to recommend high-degree items, while leaving less-popular relevant items seldom exposed.
    We should emphasize that the most popular group, \ie the 10-th group, only contains less than 1\% of item spaces (0.65\%, 0.83\%, 0.83\%, 0.22\%, respectively) but contributes more than 35\% of the total Recall scores on four datasets.
    On the contrary, the least popular group, \ie the 1-st group, contains most of the item spaces (25.51\%, 35.70\%, 31.82\%, 64.43\%, respectively) but contributes less than 3\% of the total Recall scores.
    This indicates that paired with LightGCN, BCE, BPR, and SM hardly learn high-quality representations for long-tail items.
    As such, the recommendation models suffer from popularity bias.
    What's worse, the feedback loop of recommender system will further intensify popularity bias over time, resulting in the Matthew effect~\cite{Chen2020Bias}.
    
    \item CCL and SSM perform well on long-tail groups (those with smaller group ID), showing the potential in alleviating popularity bias. This is consistent with our analysis in Section~\ref{sssec:analysis_popularity_bias}.
    Making comparisons between CCL and SSM, we find CCL performs better in the groups with the smallest group ID on three out of four datasets.
    In contrast, SSM exhibits stronger competitiveness in the waist and head groups.
    These admit that SSM balances well on normal and long-tail recommendation tasks, that is, SSM can not only promote the exposure of less popular items but also guarantee overall performance.
    Surprisingly, for SSM, the contribution of each group on Amazon-Book is nearly uniformly-distributed.
    This again justifies that, by suppressing the predicted scores of popular items, the representation learning sheds more light on long-tail items, so as to establish better representations of these items.

    % \item Both SSM loss and CCL loss show potentials in alleviating the popularity bias, which is consistent with our analyses in Section~\ref{sec:analysis_popularity_bias}.
    % % Specifically, on four datasets, the contributions of the 10-th group of SSM loss downgrade to 30.77\% 20.20\%, 26.96\% and 35.52\% respectively.
    % % On the other hand, for long-tail groups (those with smaller group ID), the Recall score of each group of SSM loss consistently outperforms other losses, which admits that the performance improvements of SSM loss mainly come from accurately recommending the less popular items.
    % On four datasets, the contributions of long-tail groups (those with smaller group ID) increase measurably. Specifically,
    % CCL loss performs best in groups with smallest group IDs on three out of four datasets.
    % However, it's less competitive in other groups, especially the waist-to-head groups.
    % Surprisingly, on Amazon-Book, the distribution of each group is nearly uniform, which again verifies that by suppressing the ratings of popular items, the representation learning sheds more lights on long-tail items, so as to establish better representations of these items.
    
    \item Another interesting observation is that: for long-tail groups, the performance rank of losses typically exhibits the following order: pointwise loss $<$ pairwise loss $<$ setwise loss~\footnote{We name the groups of SSM, SM, and CCL as the setwise loss, as they all account for a set of user-item interactions}; however, for the most popular group, the loss order reverses: setwise loss $<$ pairwise loss $<$ pointwise loss. This suggests that the limited expressiveness of binary pointwise loss and pairwise loss cannot fully capture user preference toward items. Instead, they adopt a conservative policy by recommending popular items.
\end{itemize}

\subsection{Component Analysis on SSM}
\label{ssec:exp_component_analysis}
In this section, we move on to studying different designs in SSM.
We first investigate the rationality of in-batch negative sampling strategy in SSM by making comparisons among different losses that are also equipped with in-batch negative sampling strategy. 
Then the influences of the negative sampling distribution and the number of negative samples are studied.
% We then study the influences of the negative sampling distribution and the number of negative samples.
After that, we present the impact of different similarity measurements during model training and testing. Furthermore, we conduct empirical studies on the normalization factors in message passing rules. Finally, we numerically compare the training time of different objectives.

% \wjc{In this section, we move on to studying different designs in SSM. We first investigate the rationality of in-batch negative sampling strategy in SSM by making comparisons among different losses that are also equipped with in-batch negative sampling strategy. We then study the influences of the negative sampling distribution and the number of negative samples. After that, we study the impact of different similarity measurements during model training and testing. Furthermore, we conduct empirical studies on the normalization factors in message passing rules. Lastly, we numerically compare the training time of different objectives.}

\subsubsection{\textbf{Impact of In-batch Negative Sampling Strategy}}
\label{sssec:exp_inbatch}

\begin{figure*}[t]
 \centering
 \subcaptionbox{Gowalla\label{fig:group-dis-gowalla}}{
  \includegraphics[width=0.45\textwidth]{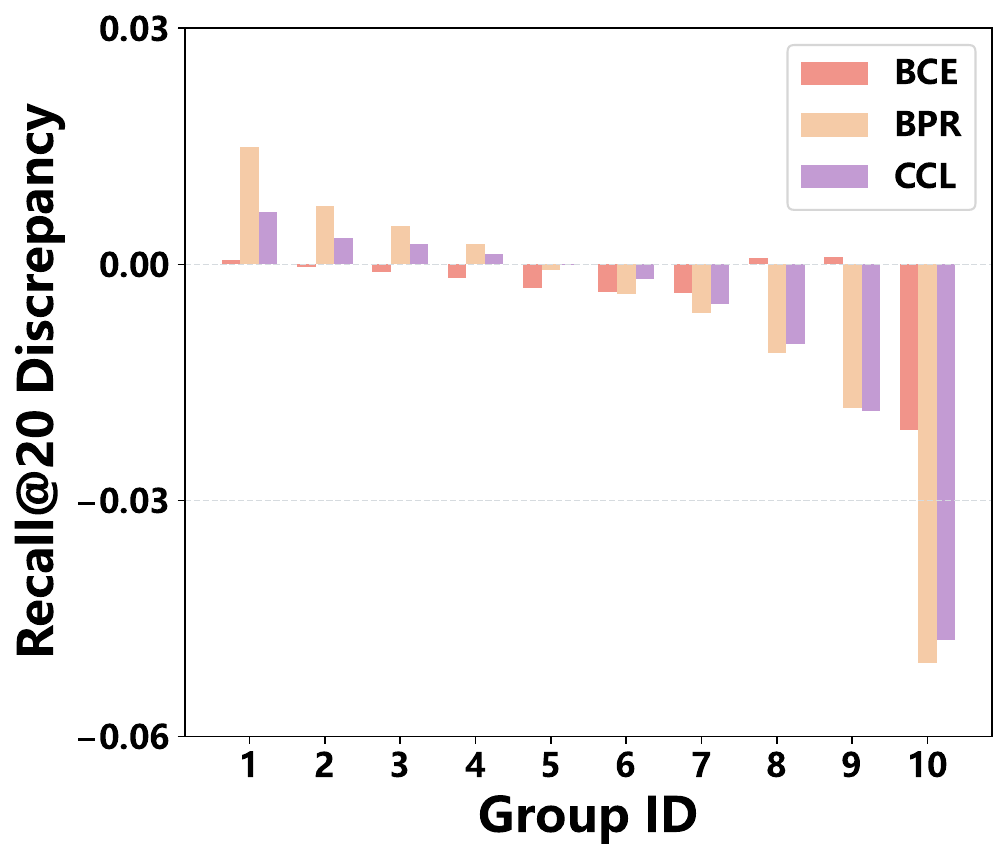}}
 \subcaptionbox{Yelp2018\label{fig:group-dis-yelp}}{
  \includegraphics[width=0.45\textwidth]{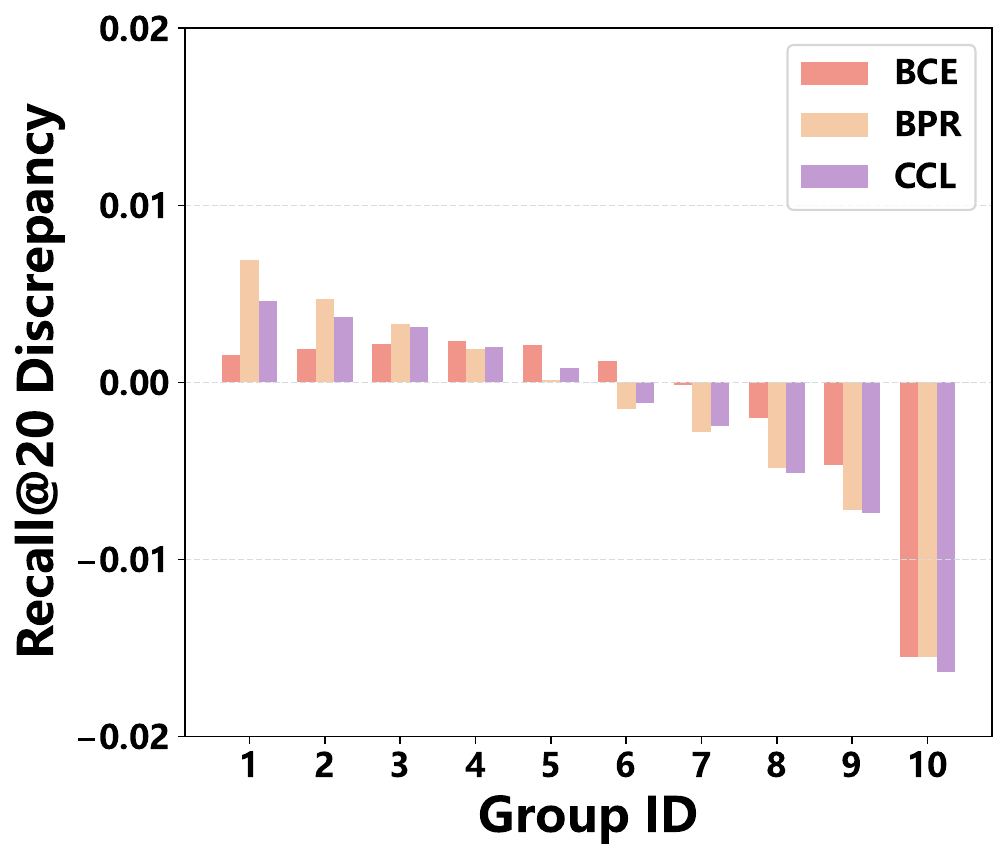}}
 \subcaptionbox{Amazon-Book\label{fig:long-tail-ab}}{
  \includegraphics[width=0.45\textwidth]{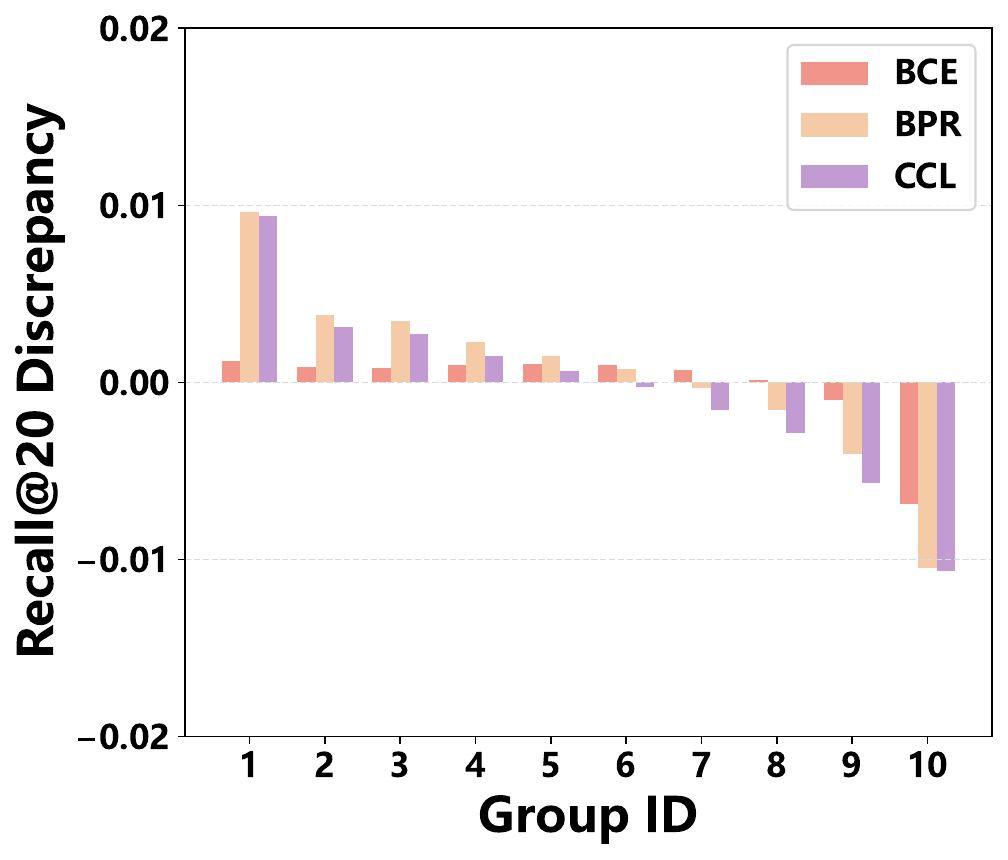}}
 \subcaptionbox{Alibaba-iFashion\label{fig:group-dis-ifashion}}{
  \includegraphics[width=0.45\textwidth]{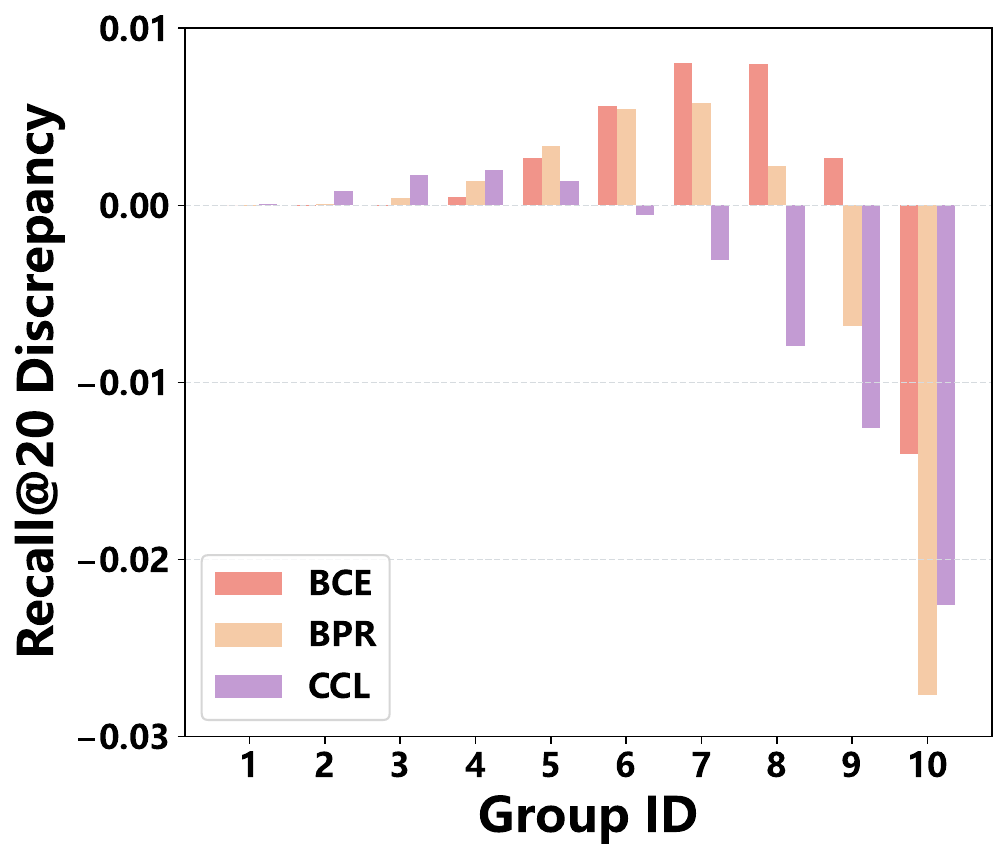}}
  % \vspace{-10pt}
  \caption{Performance discrepancy of different item groups among different objectives.}
 \label{fig:group-dis}
 % \vspace{-10pt}
\end{figure*}

By default, SSM utilizes the in-batch negative sampling strategy to make full use of parallel computing of modern hardware. Indeed, most sampling-based approaches (\eg BCE, BPR, CCL) can also be equipped with this acceleration technique. For example, we can extend BCE by assigning label `0' to other items within the same mini-batch, compared with label `1' for the corresponding positive item, termed `BCE-IB'. Similarly, we name the multiple-negative version of BPR with in-batch sharing strategy as `BPR-IB'. For CCL, instead of sampling a set of individual negative items for each observed interaction, we simply regard other items within the same mini-batch as the set of negative items, termed `CCL-IB'.
To determine the influences and the underlying mechanism of the in-batch negative sampling strategy, we compare SSM with the foregoing variants of different losses on all four datasets. Table~\ref{tab:in-batch} records the overall performance of each loss on each dataset. Fig.~\ref{fig:group-dis} shows the performance discrepancy of group-wise Recall@20 metric (\cf Equation~\eqref{eq:group-recall}) between the vanilla loss and its in-batch version.
We have the following observations:

\begin{table}[t]
\centering
\caption{Impact of the in-batch negative sampling strategy. The backbone model is LightGCN.}
\label{tab:in-batch}
% \resizebox{0.48\textwidth}{!}{
\begin{tabular}{c|cc|cc|cc|cc}
\hline
\textbf{Dataset} & \multicolumn{2}{c|}{\textbf{Gowalla}} & \multicolumn{2}{c|}{\textbf{Yelp2018}} & \multicolumn{2}{c|}{\textbf{Amazon-Book}} & \multicolumn{2}{c}{\textbf{Alibaba-iFashion}} \\ \hline
\textbf{Loss}    & \textbf{Recall}   & \textbf{NDCG}     & \textbf{Recall}    & \textbf{NDCG}     & \textbf{Recall}     & \textbf{NDCG}       & \textbf{Recall}       & \textbf{NDCG}         \\ \hline
\textbf{BCE-IB}  & 0.1429            & 0.1196            & 0.0524             & 0.0427            & 0.0361              & 0.0269              & 0.1107                & 0.0526                \\
\textbf{BPR-IB}  & 0.1225            & 0.0877            & 0.0492             & 0.0381            & 0.0470              & 0.0364              & 0.0937                & 0.0449                \\
\textbf{CCL-IB}  & 0.1093            & 0.0771            & 0.0484             & 0.0377            & 0.0495              & 0.0404              & 0.0644                & 0.0295                \\
\textbf{SSM}     & \textbf{0.1869}   & \textbf{0.1571}   & \textbf{0.0737}    & \textbf{0.0609}   & \textbf{0.0590}     & \textbf{0.0459}     & \textbf{0.1253}       & \textbf{0.0599}       \\ \hline
\end{tabular}
% }
% \vspace{-5pt}
\end{table}
% \begin{itemize}[leftmargin=*]
\begin{itemize}
    \item Jointly analyzing Table~\ref{Table:overall_comparion} and Table~\ref{tab:in-batch}, we find significant performance drop for BCE-IB, BPR-IB, and CCL-IB on all datasets, except BPR-IB on Amazon-Book. This admits that not every objective function can benefit from in-batch negative sampling strategy. SSM outperforms all compared approaches across the board, verifying the rationality and effectiveness of utilizing in-batch negative sampling strategy for SSM.
    
    \item Fig.~\ref{fig:group-dis} provides clues to explain the performance fluctuations of these compared approaches. Specifically, when equipped with in-batch negative sampling strategy, all approaches exhibit better long tail recommendation accuracy, that is, for long-tail groups (groups with smaller ID) in Fig.~\ref{fig:group-dis}, a consistent improvement \wrt recall@20 is achieved over the vanilla sampling strategy. On the contrary, the head groups suffer from serious performance deterioration. This verifies that in-batch sampling strategy has the potential to alleviate the popularity bias by over penalizing the prediction score of popular items, which is consistent with our analysis in section~\ref{sssec:analysis_popularity_bias}. However, this sampling strategy cannot guarantee the overall recommendation accuracy, which will undoubtedly limit its practical value. In contrast, as we have analyzed in section~\ref{sssec:analysis_maximizing_DCG}, SSM is highly consistent with the ranking metrics, therefore can well balance the performance of normal recommendation and long-tail recommendation.
\end{itemize}

\begin{figure}[t]
 \centering
 \subcaptionbox{Gowalla\label{fig:curve_beta_gowalla}}{
  \includegraphics[width=0.45\textwidth]{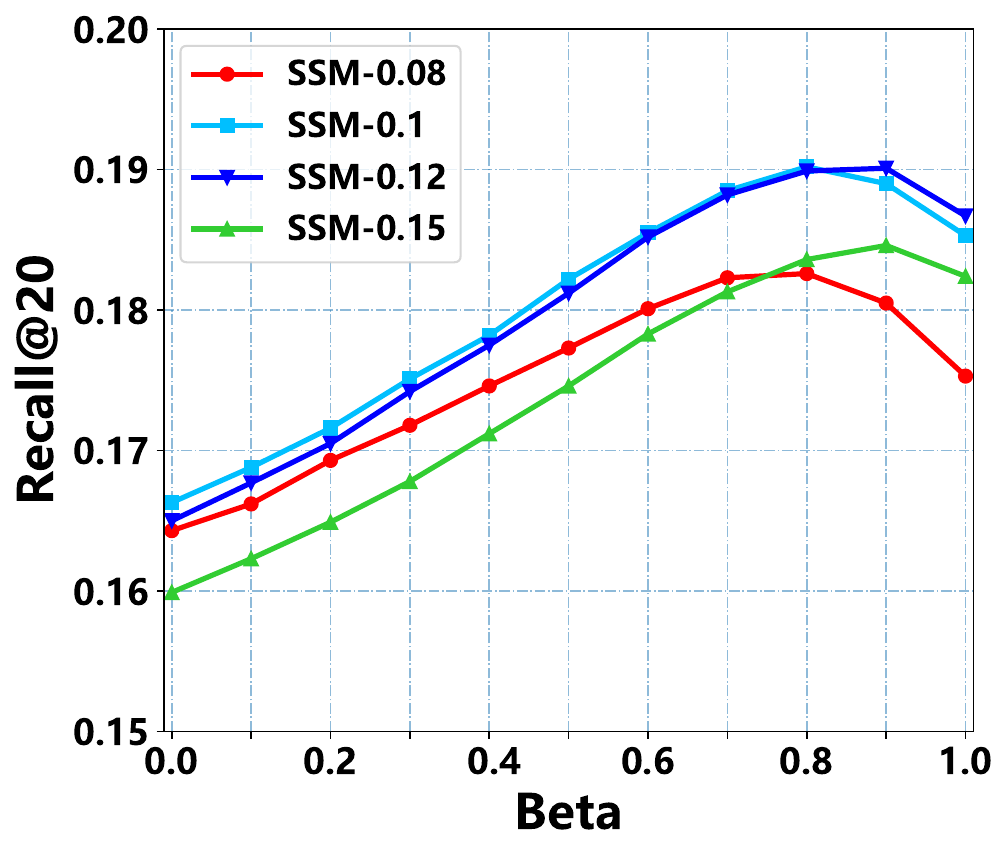}}
 \subcaptionbox{Yelp2018\label{fig:curve_beta_yelp}}{
  \includegraphics[width=0.45\textwidth]{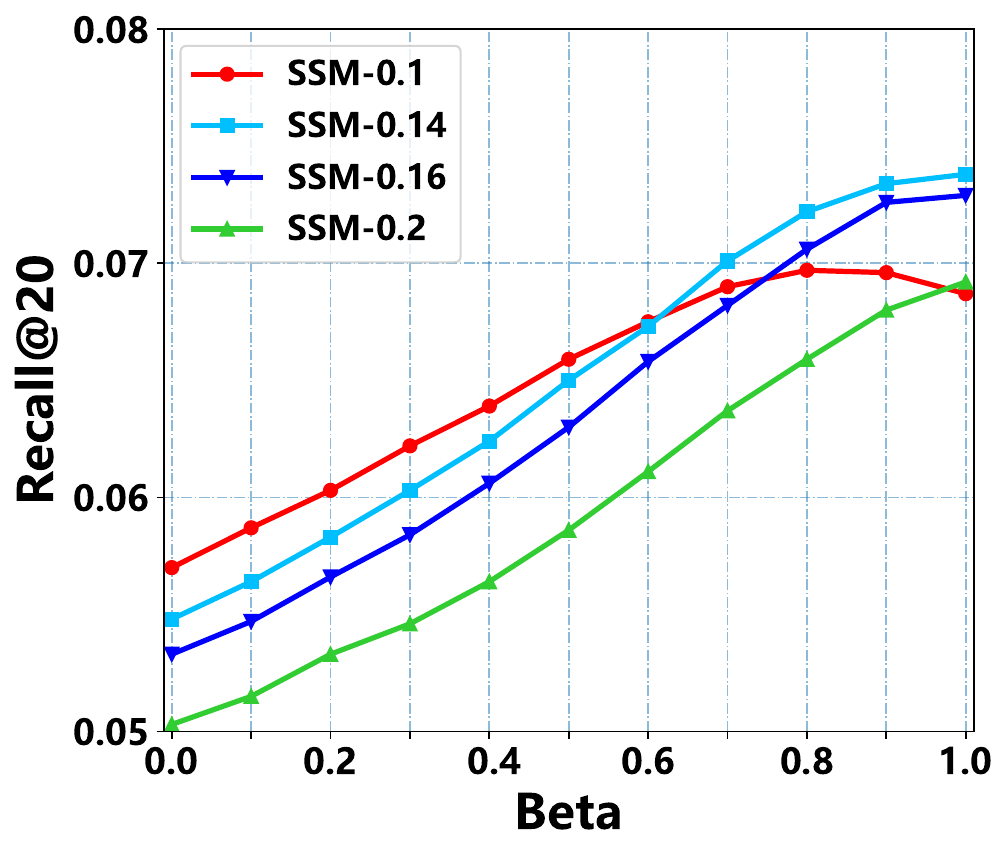}}
 \subcaptionbox{Amazon-Book\label{fig:curve_beta_ab}}{
  \includegraphics[width=0.45\textwidth]{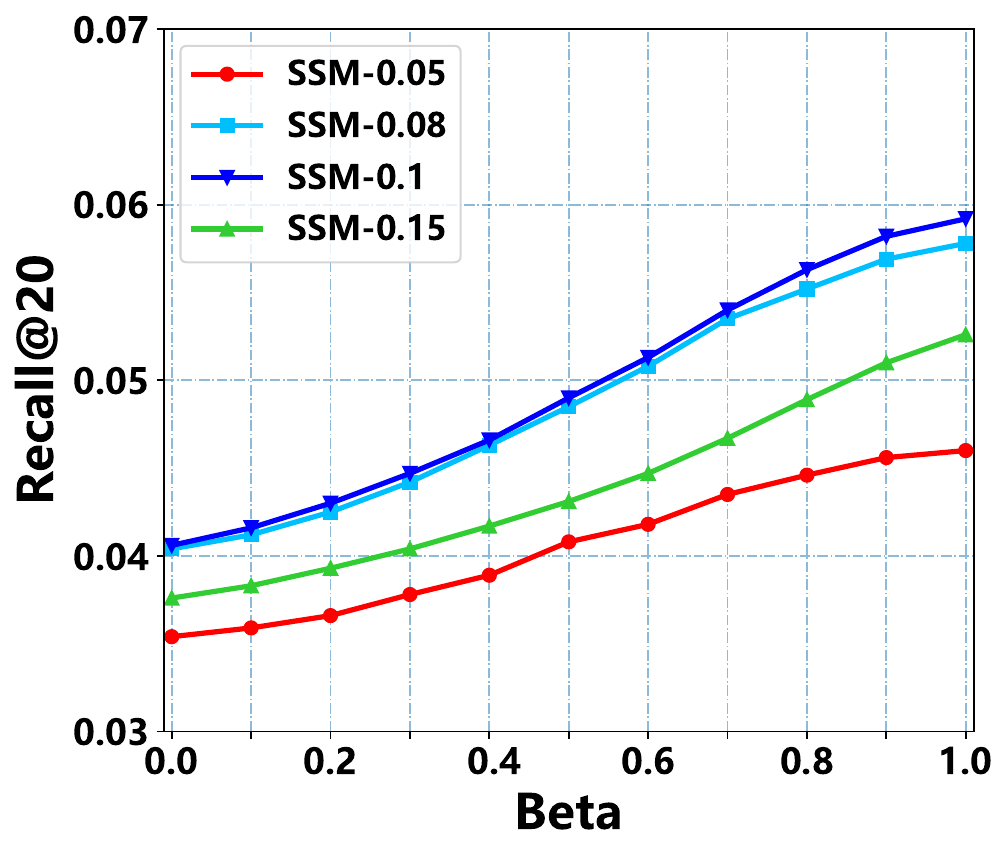}}
 \subcaptionbox{Alibaba-iFashion\label{fig:curve_beta_ifashion}}{
  \includegraphics[width=0.45\textwidth]{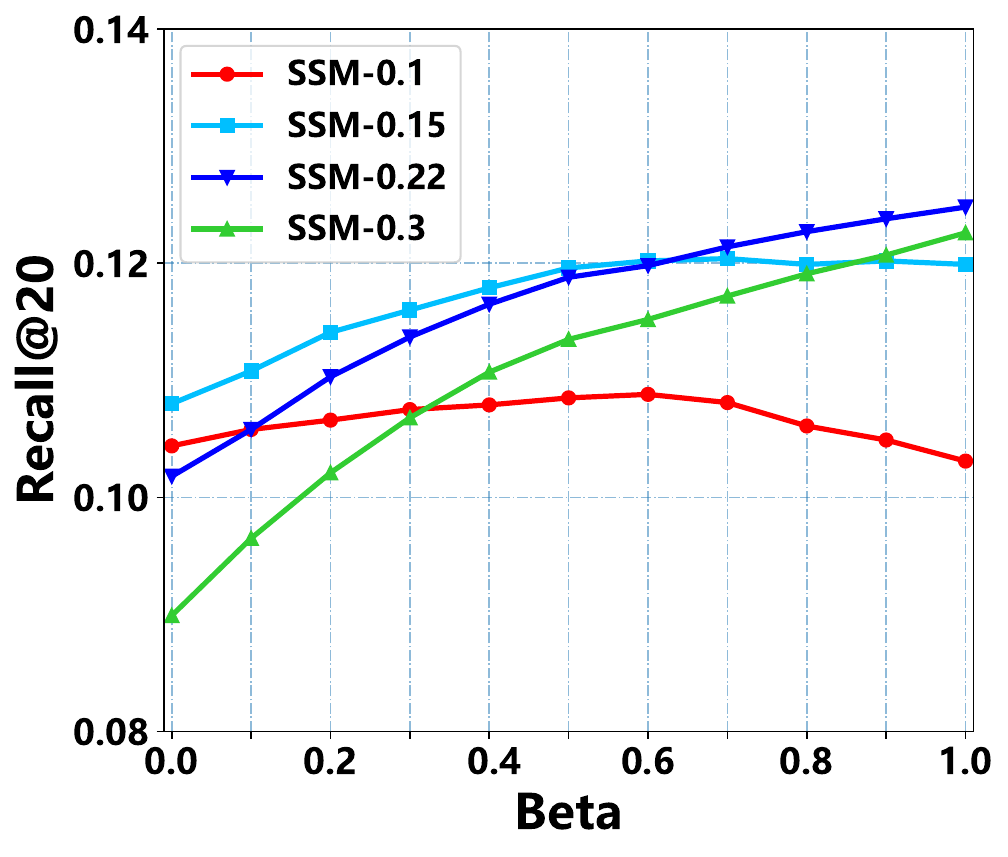}}
%   \vspace{-10pt}
  \caption{Impact of negative sampling distribution $\beta$ and temperature coefficient $\tau$.}
 \label{fig:impact-of-beta}
%  \vspace{-10pt}
\end{figure}

\subsubsection{\textbf{Impact of Negative Sampling Distribution}}
\label{sssec:exp_negative_distribution}
%%
% TODO: Add two subfigures on Yelp2018 and Amazon-Book dataset, for each subfigure, the y-axis is Recall@20 while x-axis is beta from 0 to 1, each subfigure has two curve, one is tau=OPT, the other is tau=1.
%%
As SSM is a sampling-based loss, the distribution of negative samples matters for ranking performance.
Since the default in-batch negative sampling of SSM is equivalent to sampling negative items from an empirical frequency sampler in expectation, we here introduce a variant of in-batch negative sampling to study the impact of negative sampling distribution.
Specifically, we use a customized negative sampler as the naive sampler does, but share the sampled $N$ negative samples across the positive pairs in the current batch as the in-batch negative sampling does.
For ease of implementation, we let $p_n(i)=f_i^{\beta}$, which balances between a uniform sampler and a frequency sampler, controlled by coefficient $\beta \in [0,1]$.
Other choices of $p_n$ are left as future work.

We record the fluctuation in Recall@20 as we change $\beta$ from 0 to 1 at an interval of 0.1, as shown in Fig.~\ref{fig:impact-of-beta}. We use "SSM-0.1" to represent the temperature coefficient $\tau$ of SSM is set to 0.1, and similar notations for others. We find that:

% \begin{itemize}[leftmargin=*]
\begin{itemize}
    % \item As $\beta$ increases, the ranking metric rises consistently. This admits that the negative sampling distribution indeed influences representation learning. Moreover, larger $\beta$ usually implies harder negatives since popular items are sampled with a higher probability. This also justifies the necessity of hard negative mining.
    \item With the increase in $\tau$, the ranking metric shows a rising trend in general. This admits that the negative sampling distribution indeed influences representation learning. Moreover, a larger $\beta$ usually implies harder negatives since popular items are sampled with a higher probability. This also justifies the necessity of hard negative mining.
    
    \item In most cases, the best performance is achieved at $\beta=1.0$, which supports the adoption of in-batch negative sampling strategy due to their equivalence in expectation. However, it is still possible to further improve the performance by fine-grained tuning the value of $\beta$. For example, we can obtain a better performance when setting $\beta$ to 0.9 on Gowalla, compared to the results shown in Table~\ref{Table:overall_comparion}.
    
    \item Comparing different curves, we can see that when fixing $\tau$ to a proper value (\eg 0.1 on Amazon-Book), the representation learning benefits from hard negative mining as we analyzed in Section~\ref{sssec:analysis_hard_negative_mining}. In contrast, a too large (or too small) value of $\tau$, for example, 0.15 (or 0.05) on Amazon-Book, will lead to performance degradation as it's difficult to distinguish hard negative samples from easy ones (or as few too hard negatives dominate the gradient), making the learned representation less quality.
    % The best $\tau$ typically lies in the range of $[0.10, 0.15]$
\end{itemize}

% \begin{figure}[t]
%  \centering
%  \subcaptionbox{Gowalla\label{fig:curve_beta_gowalla}}{
%   \includegraphics[width=0.45\textwidth]{figures/curve-beta-gowalla.pdf}}
%  \subcaptionbox{Yelp2018\label{fig:curve_beta_yelp}}{
%   \includegraphics[width=0.45\textwidth]{figures/curve-beta-yelp.pdf}}
%  \subcaptionbox{Amazon-Book\label{fig:curve_beta_ab}}{
%   \includegraphics[width=0.45\textwidth]{figures/curve-beta-ab.pdf}}
%  \subcaptionbox{Alibaba-iFashion\label{fig:curve_beta_ifashion}}{
%   \includegraphics[width=0.45\textwidth]{figures/curve-beta-ifashion.pdf}}
% %   \vspace{-10pt}
%   \caption{Impact of negative sampling distribution $\beta$ and temperature coefficient $\tau$.}
%  \label{fig:impact-of-beta}
% %  \vspace{-10pt}
% \end{figure}

\begin{figure}[t]
 \centering
 \subcaptionbox{Gowalla\label{fig:curve_N_gowalla}}{
  \includegraphics[width=0.45\textwidth]{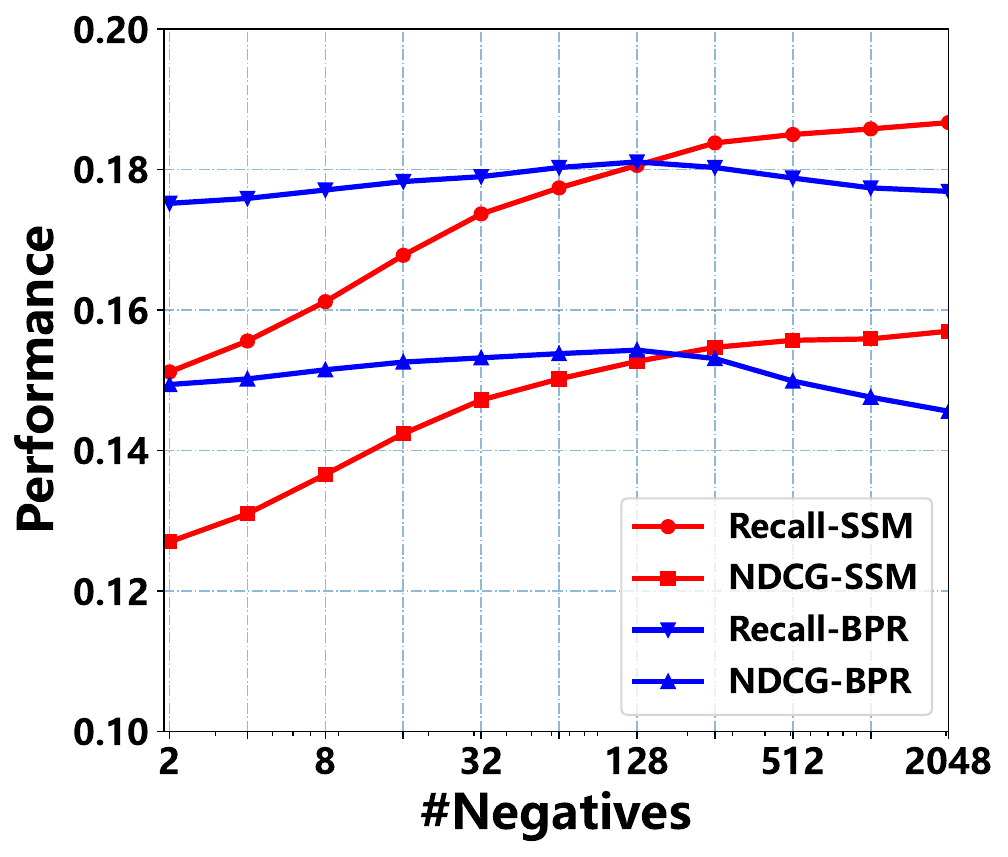}}
 \subcaptionbox{Yelp2018\label{fig:curve_N_yelp}}{
  \includegraphics[width=0.45\textwidth]{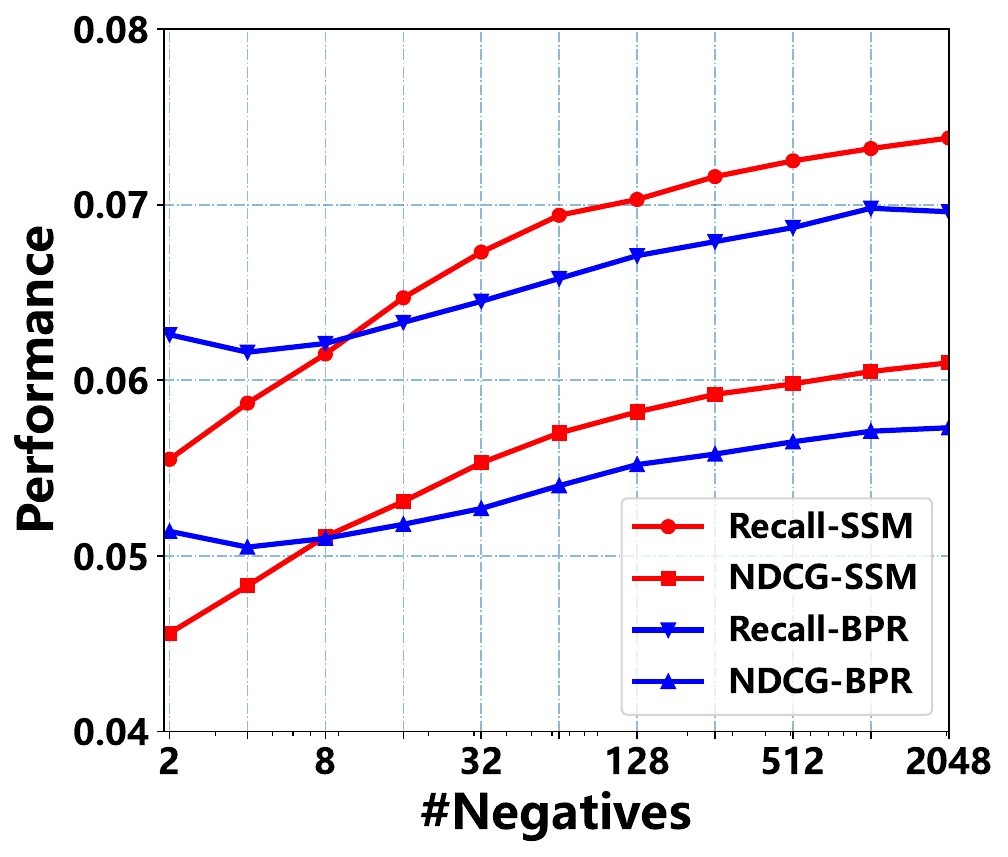}}
 \subcaptionbox{Amazon-Book\label{fig:curve_N_ab}}{
  \includegraphics[width=0.45\textwidth]{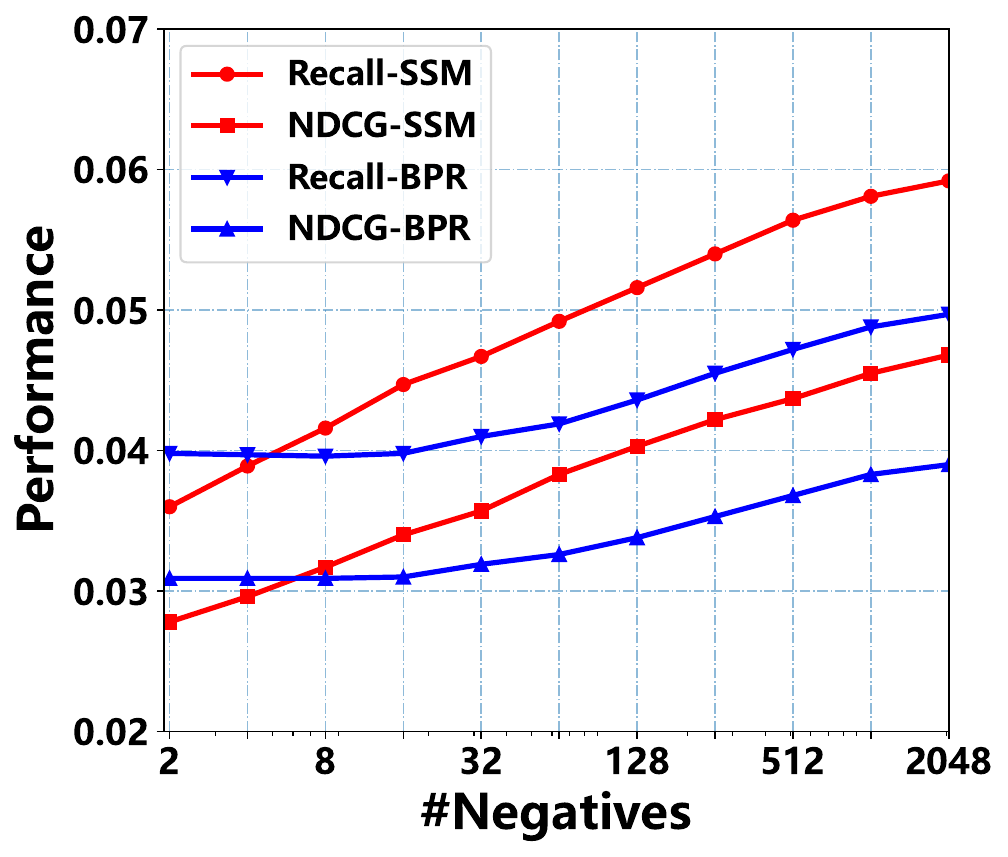}}
 \subcaptionbox{Alibaba-iFashion\label{fig:curve_N-ifahsion}}{
  \includegraphics[width=0.45\textwidth]{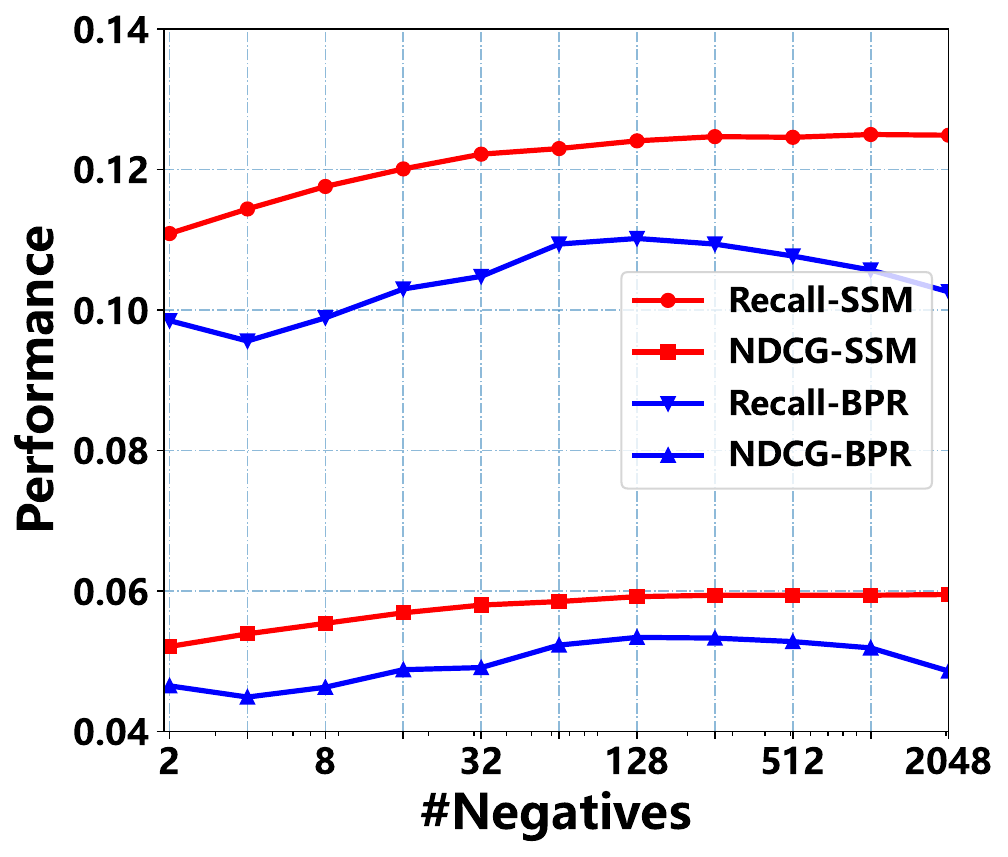}}
%   \vspace{-10pt}
  \caption{Impact of negative sampling number $N$}
 \label{fig:impact-of-N}
%  \vspace{-10pt}
\end{figure}

\subsubsection{\textbf{Impact of Negative Sampling Number}}
\label{sssec:exp_number_negatives}
%%
% TODO: Add two subfigures on Yelp2018 and Amazon-Book dataset, for each subfigure, the y-axis is Recall@20 while x-axis is the number of negatives.
%%

% \subsubsection{\textbf{Impact of Temperature Coefficient}}
% \label{sssec:exp_tau}
% %%
% % TODO: Add two subfigures on Yelp2018 and Amazon-Book dataset, for each subfigure, the y-axis is Recall@20 while x-axis is the value of tau in range 0.1 to 1.
% %%
Fig.~\ref{fig:impact-of-N} shows the impact of $N$, the number of negative samples, using the variant of SSM introduced in the previous section. Here, we extend BPR by using multiple non-interacted items as negative samples. As such, the main difference between BPR and SSM is: BPR uses inner product to measure similarity, while SSM uses temperature-aware cosine similarity. We can see that:
% \begin{itemize}[leftmargin=*]
\begin{itemize}
    \item On Yelp2018 and Amazon-Book datasets, BPR enjoys the benefits of increasing the number of negative samples. This is in line with our intuition that seeing more negative samples during model training makes a more sophisticated recommender. However, we also observe performance degeneration when $N$ exceeds some threshold, say 128 on Gowalla and Alibaba-iFashion. The possible reasons are two-fold:
    (1) increasing the size of negative samples will inevitably add difficulties to model optimization, especially when the supervision signal is highly sparse, \eg on Alibaba-iFahsion.
    (2) BPR regards all negative samples as equally important, regardless of their hardness, which leads to less-information gradients at training step~\cite{Rendle2014Improving}. Another drawback of the generalized BPR we should emphasize is its training efficiency. We can see later from Table~\ref{tab:exp_efficiency} that the generalized BPR is more than 50x slower than SSM for each training epoch when $N=2048$ on Amazon-Book.

    \item The performance of SSM keeps getting better as $N$ increases on all four datasets.
    Moreover, when $N$ becomes sufficient (\eg $N>8$ on Amazon-Book), SSM surpasses BPR, even though the total number of negative samples is $N$ in a mini-batch, in contrast to $|\mathcal{B}|\times N$ of BPR, where $|\mathcal{B}|$ is the size of mini-batch. Moreover, as $N$ increases, the performance gap becomes larger. We attribute this to the advantage of SSM in mining hard negative samples as analyzed in Section~\ref{sssec:analysis_hard_negative_mining}.
\end{itemize}

% \begin{figure}[t]
%  \centering
%  \subcaptionbox{Gowalla\label{fig:curve_N_gowalla}}{
%   \includegraphics[width=0.45\textwidth]{figures/curve-N-gowalla.pdf}}
%  \subcaptionbox{Yelp2018\label{fig:curve_N_yelp}}{
%   \includegraphics[width=0.45\textwidth]{figures/curve-N-yelp.pdf}}
%  \subcaptionbox{Amazon-Book\label{fig:curve_N_ab}}{
%   \includegraphics[width=0.45\textwidth]{figures/curve-N-ab.pdf}}
%  \subcaptionbox{Alibaba-iFashion\label{fig:curve_N-ifahsion}}{
%   \includegraphics[width=0.45\textwidth]{figures/curve-N-ifashion.pdf}}
% %   \vspace{-10pt}
%   \caption{Impact of negative sampling number $N$}
%  \label{fig:impact-of-N}
% %  \vspace{-10pt}
% \end{figure}

\subsubsection{\textbf{Impact of Different Similarity Measurements}}
\label{sssec:exp_simialrity_function}
We find an interesting phenomenon of SSM, which supports our claims about the necessity of adapting representation magnitudes during representation learning. Specifically, we use LightGCN as backbone model for SSM, and adopt different similarity measurement combinations for model training and testing --- we use either `Inner Product' similarity or `Cosine' similarity in SSM during model training, and use one of them to generate predictions during testing, resulting in four different combinations. The empirical results are shown in Table~\ref{tab:comparison_similarity_function}. We find that:

\begin{table*}[t]
\centering
\caption{Performance comparison among different similarity function combinations during training and testing phases. `IP' indicates `Inner Product' similarity while `Cos' indicates `Cosine' similarity. We report Recall@20 metric on four datasets.}
% \vspace{-10pt}
\label{tab:comparison_similarity_function}
% \begin{tabular}{c|cc|cc|cccc}
% \hline
% \multirow{2}{*}{Combination} & \multicolumn{2}{c|}{Training} & \multicolumn{2}{c|}{Testing} & \multicolumn{4}{c}{Datasets}                        \\ \cline{2-9} 
%                       & IP           & COS           & IP           & COS           & Gowalla & Yelp2018 & Amazon-Book & Alibaba-iFashion \\ \hline
% IP-IP                  & $\checkmark$ &               & $\checkmark$ &               & 0.1085  & 0.0468   & 0.0435      & 0.0749           \\
% IP-COS                 & $\checkmark$ &               &              & $\checkmark$  & 0.0825  & 0.0317   & 0.0358      & 0.0236           \\
% COS-IP                 &              & $\checkmark$  & $\checkmark$ &               & 0.1869  & 0.0737   & 0.0590      & 0.1253           \\
% COS-COS                &              & $\checkmark$  &              & $\checkmark$  & 0.1140  & 0.0477   & 0.0491      & 0.0629           \\ \hline
% \end{tabular}
\begin{tabular}{c|cc|cccc}
\hline
\multirow{5}{*}{\textbf{\begin{tabular}[c]{@{}c@{}}Similarity\\ Combination\end{tabular}}} & \multicolumn{2}{c|}{\textbf{Combination}}                    & \textbf{IP-IP}        & \textbf{IP-Cos}       & \textbf{Cos-IP}       & \textbf{Cos-Cos}      \\ \cline{2-7} 
                                                                                           & \multirow{2}{*}{\textbf{training}}         & \textbf{IP}     & \textbf{$\checkmark$} & \textbf{$\checkmark$} & \textbf{}             & \textbf{}             \\
                                                                                           &                                            & \textbf{Cos}    & \textbf{}             & \textbf{}             & \textbf{$\checkmark$} & \textbf{$\checkmark$} \\ \cline{2-7} 
                                                                                           & \multirow{2}{*}{\textbf{testing}}          & \textbf{IP}     & \textbf{$\checkmark$} & \textbf{}             & \textbf{$\checkmark$} & \textbf{}             \\
                                                                                           &                                            & \textbf{Cos}    & \textbf{}             & \textbf{$\checkmark$} & \textbf{}             & \textbf{$\checkmark$} \\ \hline
\multirow{8}{*}{\textbf{Performance}}                                                          & \multirow{2}{*}{\textbf{Gowalla}}          & \textbf{Recall} & 0.1085                & 0.0825                & \textbf{0.1867}       & 0.1140                \\
                                                                                           &                                            & \textbf{NDCG}   & 0.0768                & 0.0551                & \textbf{0.1567}       & 0.0794                \\ \cline{2-7} 
                                                                                           & \multirow{2}{*}{\textbf{Yelp2018}}         & \textbf{Recall} & 0.0468                & 0.0317                & \textbf{0.0737}       & 0.0477                \\
                                                                                           &                                            & \textbf{NDCG}   & 0.0369                & 0.0241                & \textbf{0.0609}       & 0.0371                \\ \cline{2-7} 
                                                                                           & \multirow{2}{*}{\textbf{Amazon-Book}}      & \textbf{Recall} & 0.0435                & 0.0358                & \textbf{0.0590}       & 0.0491                \\
                                                                                           &                                            & \textbf{NDCG}   & 0.0336                & 0.0272                & \textbf{0.0459}       & 0.0381                \\ \cline{2-7} 
                                                                                           & \multirow{2}{*}{\textbf{Alibaba-iFashion}} & \textbf{Recall} & 0.0749                & 0.0236                & \textbf{0.1253}       & 0.0629                \\
                                                                                           &                                            & \textbf{NDCG}   & 0.0357                & 0.0095                & \textbf{0.0599}       & 0.0269                \\ \hline
\end{tabular}
\end{table*}

% \begin{itemize}[leftmargin=*]
\begin{itemize}
    \item Cosine similarity is a better choice than Inner Product similarity for SSM during training.
    This is consistent with our analysis in Section~\ref{sssec:analysis_hard_negative_mining}: equipped with temperature-aware cosine similarity, SSM is capable of performing hard negative mining, so as to enhance the quality of the learned representations.
    A nice property of Cosine similarity is that the similarity value is bounded in the interval $[-1,1]$, making it easier to train.
    \item Compared with Cosine similarity, Inner Product similarity performs better for SSM during testing.
    The underlying reason is that in addition to the angle between user and item representations, Inner Product similarity additionally considers the magnitude of two representations, thus making full use of the learned representations for prediction.
\end{itemize}

\subsubsection{\textbf{Empirical study on magnitude of item representations}}
\label{sssec:exp_magnitude}

\begin{figure}[t]
 \centering
 \subcaptionbox{Gowalla\label{fig:curve_mag_gowalla}}{
  \includegraphics[width=0.45\textwidth]{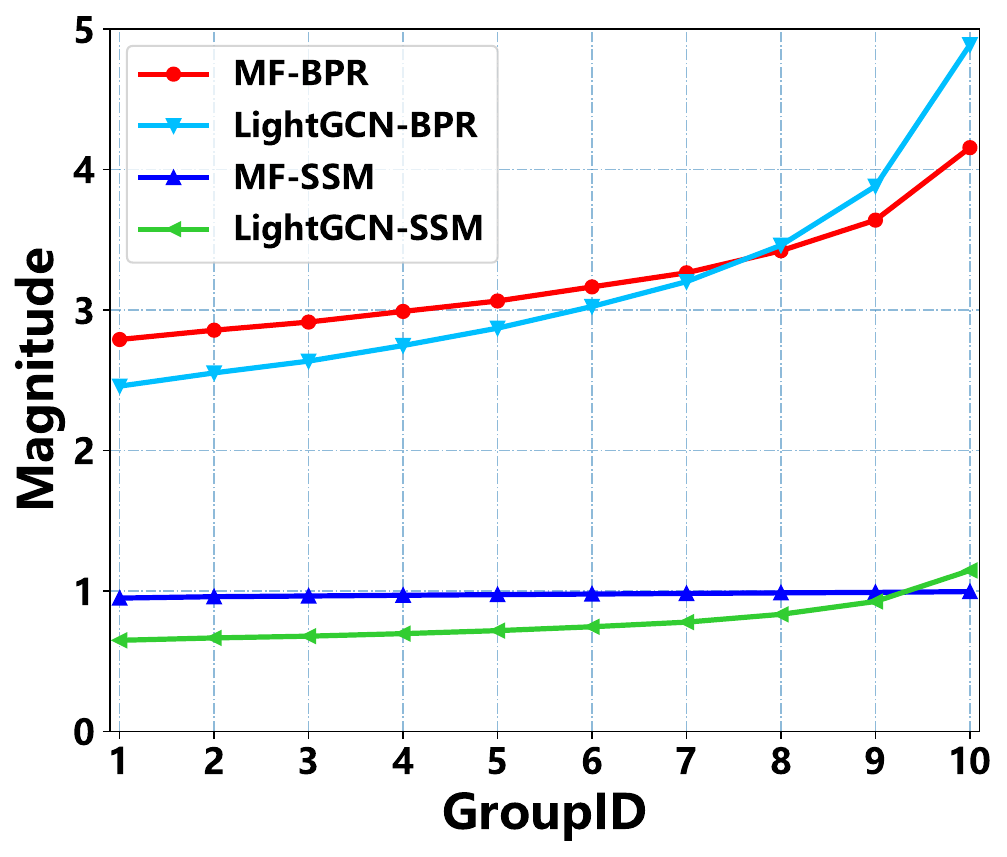}}
 \subcaptionbox{Yelp2018\label{fig:curve_mag_yelp}}{
  \includegraphics[width=0.45\textwidth]{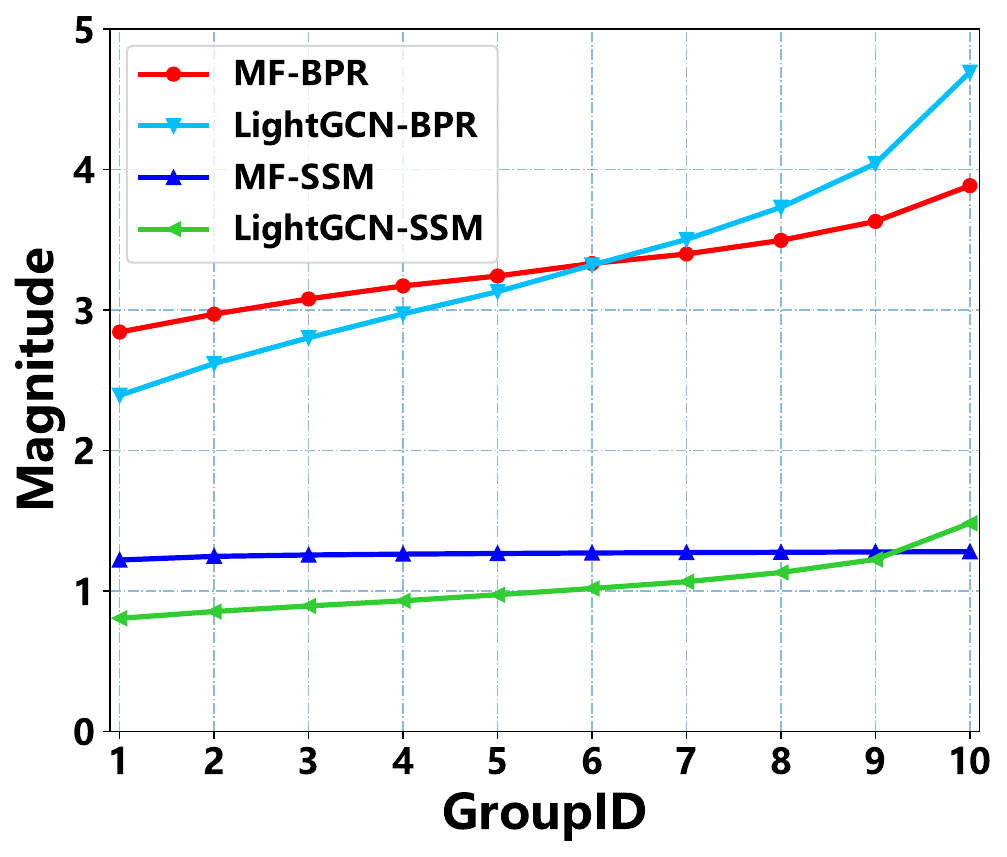}}
 \subcaptionbox{Amazon-Book\label{fig:curve_mag_ab}}{
  \includegraphics[width=0.45\textwidth]{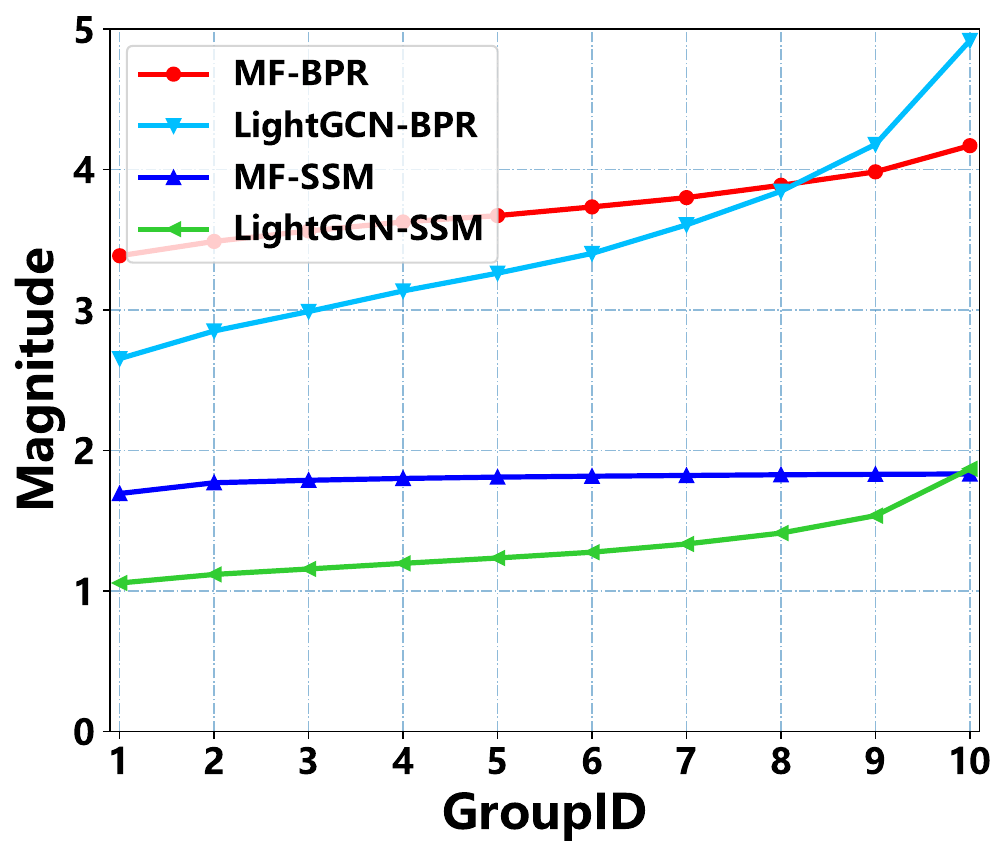}}
 \subcaptionbox{Alibaba-iFashion\label{fig:curve_mag_ifashion}}{
  \includegraphics[width=0.45\textwidth]{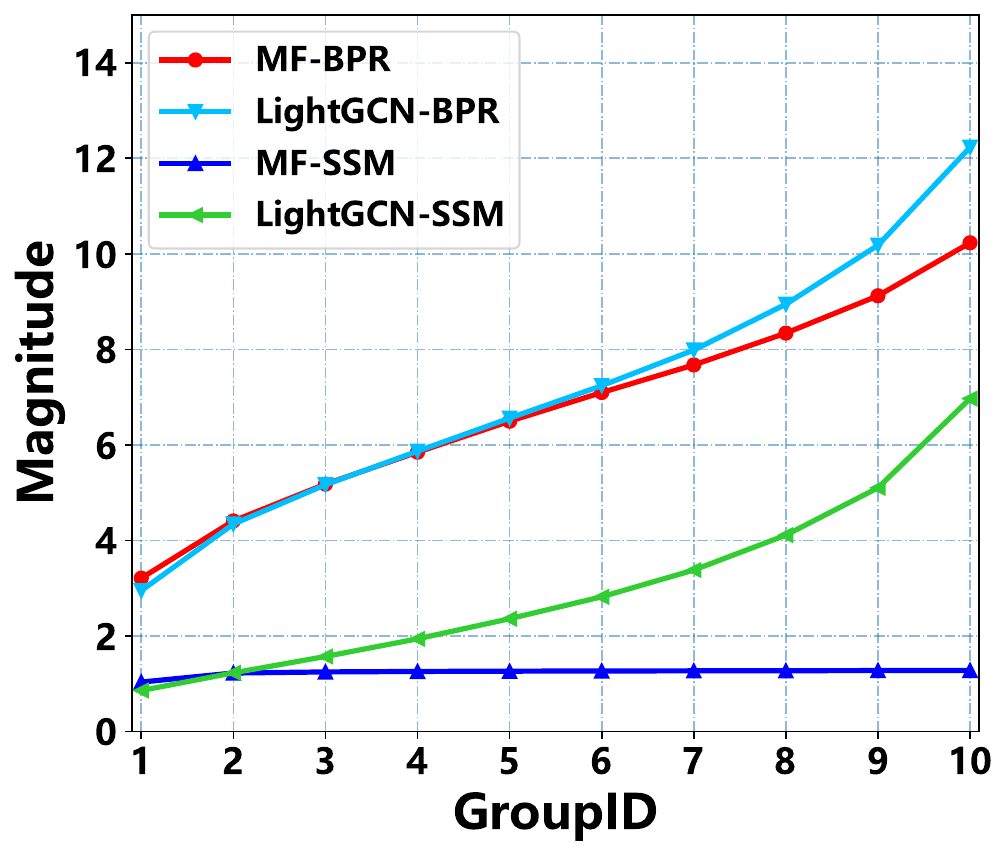}}
%   \vspace{-10pt}
  \caption{Empirical study on the magnitude of item representations learned by BPR and SSM.}
 \label{fig:empirical_magnitude}
%  \vspace{-10pt}
\end{figure}
By default, SSM uses the cosine function to measure user-item similarity, which implies that item magnitudes do not contribute to the loss and cannot guide magnitude learning. To provide empirical evidence for this claim, we conducted a study comparing the item representation magnitudes learned by SSM and BPR, using MF and LightGCN as backbones, creating four models: `MF-SSM', `LightGCN-SSM', `MF-BPR', and `LightGCN-BPR'. To ensure fair comparisons and avoid the impact caused by the L2 regularization term, we removed it from the objective function and fixed the number of epochs to 50. We split items into ten groups based on frequency while keeping the
total number of interactions of each group the same, then record the average magnitude of items within each group.
Figure~\ref{fig:empirical_magnitude} shows the results on different datasets. We can observe that:
\begin{itemize}
    \item When using MF as the backbone, the curve of item magnitude learned by SSM within different groups is flat, which confirms that SSM cannot adjust the item magnitude. However, graph-based models (\eg LightGCN) can adjust item magnitude itself, compensating for the shortcoming of SSM and achieving excellent performance (\cf Table~\ref{Table:overall_comparion}).
    \item BPR uses inner product as the similarity function, which allows it to adjust the magnitude of items during training. However, this can lead to popularity bias, especially for graph-based models such as LightGCN-BPR, where popular items typically obtain much larger magnitudes and hence have larger similarity scores. This makes popular items even more popular and results in suboptimal performance.
\end{itemize}

\subsubsection{\textbf{Empirical study on message passing strategy}}
\label{sssec:exp_message_passing}
As analyzed in Section~\ref{sssec:analysis_message_passing}, the normalization factors $\alpha_0$ and $\alpha_1$, which determine the message passing rule, play important roles in adjusting the magnitude of learned representations.
To see how $\alpha_0$ and $\alpha_1$ influence the ranking performance, we adopt SVD++ which can be viewed as the simplest message-passing-based method as the backbone model, and train it using SSM.
Note that the vanilla SVD++ only propagates messages on the user side, here we introduce a variant that only propagates messages on the item side.
Similar to the formulae of vanilla SVD++ on the user side, we define the propagation rule of SVD++ on the item side as follows:
\begin{gather}
\label{eq:item_SVD++}
    \Mat{p}_u = \Mat{p}_u^{(0)}, \quad \Mat{q}_i = \Mat{q}_i^{(0)} + \sum_{u\in \mathcal{P}_i} \frac{1}{|\mathcal{P}_i|^{\alpha_0} |\mathcal{P}_u|^{\alpha_1}} \Mat{p}_u^{(0)},
\end{gather}
where $\Mat{p}_u^{(0)}$ (or $\Mat{q}_i^{(0)}$) and $\Mat{p}_u$ (or $\Mat{q}_i$) are the ID embedding and final representation of user $u$ (or item $i$), respectively.
Table~\ref{tab:message_passing} shows the Recall@20 and NDCG@20 metrics on all four datasets.
% are shown in Table~\ref{tab:message_passing}.
We can observe that:
% \begin{itemize}[leftmargin=*]
\begin{itemize}
    \item With different values of $\alpha_0$ and $\alpha_1$, the ranking performance varies considerably.
    Typically, compared with $\alpha_1$, changing the value of $\alpha_0$ will cause larger performance fluctuations.
    % Typically, the impact of $\alpha_0$ is larger than $\alpha_1$.
    This is in line with our analyses that $\alpha_0$ determines the order of the representation magnitude while $\alpha_1$ influences its multiplication factor.
    \item The best choices of $\alpha_0$ and $\alpha_1$ are different in user-side SVD++ and item-side SVD++.
    More specifically, on all datasets, the best performance of user-side SVD++ is achieved when $\alpha_0=1.0$ and $\alpha_1=0$. While for item-side SVD++, $\alpha_0=\alpha_1=0.5$ is the best option across four datasets.
    \item On Gowalla, Yelp2018, and Amazon-Book datasets, item-side SVD++ can obtain better performance than user-side SVD++, which is in contrast to the case on Alibaba-iFashion.
\end{itemize}

% \begin{table}[h]
% \label{Table:comparison_sampled_softmax}
% \resizebox{0.48\textwidth}{!}{
% \begin{tabular}{c|cc|cc|cc|cc}
% \hline
% Dataset       & \multicolumn{2}{c|}{Gowalla} & \multicolumn{2}{c|}{Yelp2018} & \multicolumn{2}{c|}{Amazon-Book} & \multicolumn{2}{c}{Alibaba-iFahsion} \\ \hline
% {[}$\alpha_0,\alpha_1${]}        & User          & Item         & User          & Item          & User            & Item           & User              & Item             \\ \hline
% {[}0.5,0{]}   & xxxxxx        & xxxxxx       & 0.0446        & 0.0682        & 0.0466          & 0.0463         & 0.1212            & 0.1202           \\ \hline
% {[}0.5,0.5{]} & xxxxxx        & xxxxxx       & 0.0503        & 0.0730        & 0.0489          & 0.0560         & 0.1142            & 0.1213           \\ \hline
% {[}1.0,0{]}   & xxxxxx        & xxxxxx       & 0.0685        & 0.0408        & 0.0543          & 0.0455         & 0.1272            & 0.0582           \\ \hline
% \end{tabular}}
% \end{table}

\begin{table}[t]
\caption{Impact of message passing strategy. `User' indicates propagating messages on the user side, while `Item' indicates propagating messages on the item side.}
\label{tab:message_passing}
\resizebox{\textwidth}{!}{
% \begin{tabular}{c|cc|cc|cc}
% \hline
% Dataset       & \multicolumn{2}{c|}{Yelp2018} & \multicolumn{2}{c|}{Amazon-Book} & \multicolumn{2}{c}{Alibaba-iFahsion} \\ \hline
% {[}$\alpha_0,\alpha_1${]}        & User          & Item          & User            & Item           & User              & Item             \\ \hline \hline
% {[}0.5,0{]}   & 0.0446        & 0.0682        & 0.0466          & 0.0463         & 0.1212            & 0.1202           \\ \hline
% {[}0.5,0.5{]} & 0.0503        & 0.0730        & 0.0489          & 0.0560         & 0.1142            & 0.1213           \\ \hline
% {[}1.0,0{]}   & 0.0693        & 0.0408        & 0.0547          & 0.0455         & 0.1276            & 0.0582           \\ \hline
% \end{tabular}
\begin{tabular}{ccc|ccc|ccc}
\hline
\multicolumn{3}{c|}{\textbf{Propagation Side}}                                                                            & \multicolumn{3}{c|}{\textbf{User}}                                    & \multicolumn{3}{c}{\textbf{Item}}                                    \\ \hline
\multicolumn{3}{c|}{\textbf{Setting of {[}$\alpha_0,\alpha_1${]}}}                                                        & \textbf{{[}0.5,0{]}} & \textbf{{[}0.5,0.5{]}} & \textbf{{[}1.0,0{]}}  & \textbf{{[}0.5,0{]}} & \textbf{{[}0.5,0.5{]}} & \textbf{{[}1.0,0{]}} \\ \hline
\multicolumn{1}{c|}{\multirow{8}{*}{\textbf{Performance}}} & \multirow{2}{*}{\textbf{Gowalla}}          & \textbf{Recall} & 0.1156               & 0.1184                 & \textbf{0.1672}       & 0.1746               & \textbf{0.1875$^{*}$}  & 0.0969               \\
\multicolumn{1}{c|}{}                                      &                                            & \textbf{NDCG}   & 0.0855               & 0.0828                 & \textbf{0.1373}       & 0.1461               & \textbf{0.1567$^{*}$}  & 0.0656               \\ \cline{2-9} 
\multicolumn{1}{c|}{}                                      & \multirow{2}{*}{\textbf{Yelp2018}}         & \textbf{Recall} & 0.0446               & 0.0503                 & \textbf{0.0685}       & 0.0682               & \textbf{0.0737$^{*}$}  & 0.0408               \\
\multicolumn{1}{c|}{}                                      &                                            & \textbf{NDCG}   & 0.0349               & 0.0392                 & \textbf{0.0565}       & 0.0565               & \textbf{0.0611$^{*}$}  & 0.0311               \\ \cline{2-9} 
\multicolumn{1}{c|}{}                                      & \multirow{2}{*}{\textbf{Amazon-Book}}      & \textbf{Recall} & 0.0466               & 0.0489                 & \textbf{0.0543}       & 0.0463               & \textbf{0.0560$^{*}$}  & 0.0455               \\
\multicolumn{1}{c|}{}                                      &                                            & \textbf{NDCG}   & 0.0366               & 0.0382                 & \textbf{0.0427}       & 0.0365               & \textbf{0.0440$^{*}$}  & 0.0360                \\ \cline{2-9} 
\multicolumn{1}{c|}{}                                      & \multirow{2}{*}{\textbf{Alibaba-iFashion}} & \textbf{Recall} & 0.1212               & 0.1050                 & \textbf{0.1272$^{*}$} & 0.1202               & \textbf{0.1213}        & 0.0596               \\
\multicolumn{1}{c|}{}                                      &                                            & \textbf{NDCG}   & 0.0592               & 0.0505                 & \textbf{0.0622$^{*}$} & 0.0574               & \textbf{0.0577}        & 0.0261               \\ \hline
\end{tabular}
}
% \vspace{-5pt}
\end{table}

\subsubsection{\textbf{Efficiency Comparison}}
\label{sssec:exp_efficiency}
%%
% TODO: Add two subfigures on Amazon-Book dataset, one is training time per epoch as n_negs increases, the other is Recall@20 vs. epoch
%%
\begin{table}[h]
\centering
\caption{Efficiency comparison on Amazon-Books using a three-layer LightGCN as backbone model. $N$ is the number of negative samples.}
% \vspace{-10pt}
\label{tab:exp_efficiency}
% \resizebox{0.46\textwidth}{!}{
\begin{tabular}{c|c|c|c}
\hline
Loss        & Time/Epoch & Best Epoch & Total Time \\ \hline
BCE ($N$=4)    & 268s       & 61         & 4h32m    \\ \hline
BPR ($N$=1)    & 48s        & 700        & 9h20m    \\ \hline
BPR ($N$=2048) & 1983s      & 45         & 24h47m   \\ \hline
SM          & 81s        & 29         & 39m      \\ \hline
CCL ($N$=2048) & 124s       & 34         & 1h10m    \\ \hline
SSM ($N$=2048) & 39s        & 21         & 13m      \\ \hline
\end{tabular}
% }
\end{table}

We study the training efficiency of SSM. Specifically, we conduct experiments on the same Nvidia Titan RTX graphics card equipped with an Inter i7-9700K CPU (32GB Memory). We compare them under the same implementation framework based on TensorFlow, using the same acceleration methods (\ie accelerating the sampling with C++) to ensure fairness. The backbone model is a three-layer LightGCN.
The results are reported in Table~\ref{tab:exp_efficiency}. We can find that equipped with in-batch negative sampling and temperature-aware cosine similarity, SSM is much more efficient than other baselines in both the average training time per epoch and the number of epochs to reach the best performance, meanwhile achieving the leading performance (\cf Table~\ref{Table:overall_comparion}).
Surprisingly, SM achieves the second-best training efficiency on the Amazon-Book dataset. However, as mentioned previously, computing SM can be computationally expensive for large datasets since it involves exponentiating all the scores of candidate items. To verify this, we enriched the candidate item pool with a preset number (\#Paddings in Table~\ref{tab:exp_memory}) while keeping the training and testing samples unchanged. We then recorded the GPU memory and time cost per epoch of SM and SSM using a three-layer LightGCN as the backbone model on the Amazon-Book dataset, as shown in Table~\ref{tab:exp_memory}. Our results demonstrate that the time cost of SM increases linearly as the number of padding items increases. When the padding number reaches 500,000, SM suffers from out-of-memory (OOM) errors. In contrast, SSM's memory cost is low, and the running time remains almost unchanged across different numbers of padding items. These findings indicate that SSM is more efficient than SM in terms of memory usage and may be a better choice when dealing with real-world large-scale recommender systems.

\begin{table}[h]
\centering
\caption{Memory cost comparison on Amazon-Books using a three-layer LightGCN as backbone model.}
% \vspace{-10pt}
\label{tab:exp_memory}
% \resizebox{0.46\textwidth}{!}{
\begin{tabular}{c|cc|cc}
\hline
\multirow{2}{*}{\#Paddings} & \multicolumn{2}{c|}{SM} & \multicolumn{2}{c}{SSM} \\
                            & GPU Memory & Time/Epoch & GPU Memory & Time/Epoch \\ \hline
0                           & 4865M      & 81s        & 1793M      & 39s        \\
100,000                     & 5873M      & 132s       & 1793M      & 41s        \\
200,000                     & 9973M      & 183s       & 2801M      & 42s        \\
300,000                     & 9973M      & 231s       & 2801M      & 43s        \\
400,000                     & 10993M     & 280s       & 2801M      & 44s        \\
500,000                     & OOM        & -          & 2801M      & 45s        \\ \hline
\end{tabular}
% }
\end{table}
% !TeX root = ./0_main.tex
\section{Related Work}
In this section, we review the popular objective functions for item recommendation, which cast the task into a supervised learning problem.

%% todo: pointwise loss
Many traditional item recommendation methods perform learning by minimizing the \textbf{pointwise} divergence of the reconstructed interaction matrix from the observed interaction matrix.
According to the way the divergence is defined, pointwise losses can be further categorized into mean square error loss~\cite{Hu2008Collaborative,Yun2014NOMAD,Ning2011SLIM,Koren2008Factorization}, binary cross-entropy loss~\cite{He2017NCF,Shan2016Deep}, and hinge loss~\cite{Rennie2005Fast}, to name a few.
%% todo: pairwise loss
However, for item recommendation, the primary goal is not to score the interaction exactly to 1 or 0.
Instead, the relative ordering of a positive item over an unobserved item is of central importance.
Toward this goal, \textbf{pairwise} losses are proposed to optimize the preference structure consistency between the reconstructed interaction matrix from the observed interaction matrix.
Specifically, pairwise losses treat training data as a set of triplet instances $\{(u,i,j)\}$ to capture the intention that user $u$ prefers to positive item $i$ over irrelevant item $j$.
BPR~\cite{Rendle2009BPR} is one of the most popular pairwise losses in item recommendation which is proven to optimize AUC scores.
WARP~\cite{Weston2011WSABIE} is another pairwise algorithm customized for item recommendation. It encourages the predicted score of positive item larger than that of negative item above a margin, which is in line with the hinge loss.
Similarly, \cite{Park2015Preference} proposed a max-margin hinge ranking loss to minimize the ranking risk in the reconstructed matrix.
\cite{Kabbur2013FISM} devised a pairwise MSE loss that computes the relative difference between the actual non-zero and zero entries and the difference between their corresponding predicted entries.
%% todo: softmax loss and sampled softmax loss, importance sampling, contrastive loss

Besides pointwise losses and pairwise losses, a third option for item recommendation is to model the recommendation predictions over all items into a probability distribution that is normalized using a softmax function, termed the \textbf{softmax} loss~\cite{Rendle2021Item}. It's worth mentioning that in the community of Learning-to-Ranking, this type of loss function is generally termed list-wise loss~\cite{CaoQLTL07Learning,XiaLWZL08Listwise}. We will not go deep into it since it is not within the scope of this work.
Prior work~\cite{Bruch2019An} verified that softmax loss aligns well with the ranking metrics.
However, calculating the partition function of softmax is computationally costly.
A substitute for softmax loss is SSM loss which reduces the computational cost by employing the partition function only on a small subset of negatives.
However, common view on SSM loss is that it's a biased version of full softmax loss which can be corrected by log$Q$ correction~\cite{Bengio2003Quick}.
Agreeing with this point of view, some follow-on works in recommendation devise different methods to get unbiased estimation\cite{Wang2021Cross,Yang2020Mixed,Yi2019Sampling,Bai2017TAPAS,Blanc2018Adaptive}.
For example, \citet{Blanc2018Adaptive} proposed a divide and conquer algorithm to estimate the proposal distribution with kernel based sampling strategy.
\citet{Yi2019Sampling} presented a method to estimate item frequency for streaming data without requiring fixed item vocabulary.
\citet{Wang2021Cross} devised a cross-batch negative sampling with the FIFO memory bank to improve the accuracy of estimating item distribution by enlarging the number of negative samples.
% enlarge the number of negative samples, improving the accuracy of estimating item distribution.

Most recently, a surge of works introduced contrastive loss like InfoNCE loss into recommendation~\cite{Wu2021Self,Zhou2021Contrastive,Tang2021Multi,Mao2021SimpleX,Zhang22BC}.
At its core is maximizing the agreement of positive pairs as compared with that of negative pairs.
% They typically used contrastive loss as the auxiliary task's loss for better representation learning~\cite{Wu2021Self,Pan2021Efficient}.
Interestingly, when directly utilizing the supervision signal (whether the interaction is observed) of the data to construct positive and negative pairs, InfoNCE loss becomes SSM loss.
However, to the best of our knowledge, only very limited works~\cite{Zhou2021Contrastive,Tang2021Multi,Liu2021Contrastive} utilized SSM loss as their main task objective to train recommendation model.
\citet{Zhou2021Contrastive} proved that both contrastive loss and multinomial IPW loss optimize the same objective in principle but failed to answer why contrastive loss works.
\citet{Liu2021Contrastive} presented a debiased contrastive loss to remedy the negative effects of false negative samples in the randomly sampled subset.
MSCL \cite{Tang2021Multi} is mostly related to our work which also realized the superiority of using SSM loss to train recommender.
However, it differs from our work in:
(1) MSCL \cite{Tang2021Multi} only empirically verified the effectiveness of SSM loss for graph-based recommender. In contrast, we theoretically reveal three advantages of SSM loss, that is, mitigating popularity bias, mining hard negative samples, and maximizing the ranking metrics;
(2) We also uncover the shortcomings of SSM loss in adjusting the representation magnitudes and present theoretical evidence to support our argument. 
% !TeX root = ./0_main.tex
\section{conclusion and future work}
In this work, we present insightful analyses of SSM for item recommendation.
Firstly, we theoretically disclose model-agnostic advantages of SSM in mitigating popularity bias, mining hard negative samples and maximizing ranking metrics.
We then probe the model-specific characteristics of SSM and point out its potential shortcoming in adjusting representation magnitudes.
To justify our argument, we further show that message passing based methods are capable of adjusting magnitude.
We conducted extensive experiments on four benchmark datasets, demonstrating the superiority of training history- and graph-based models using SSM for both normal and long tail recommendation tasks.

We believe the comprehending of SSM is inspirational to future developments of recommendation community.
In future work, we would like to further refine and augment the capabilities of SSM.
A key area of focus will be enhancing its robustness through distributionally robust optimization \cite{Yang0ZW0CW23DROS,wu2023understanding}.
We also plan to explore the applicability of SSM to a wider range of model architectures, such as the diffusion model \cite{Yang23DreamRec}.
Moreover, since practical recommender systems usually involve rich side information, exploring the potential of SSM for feature-based CTR models is another promising direction.

%%
%% The acknowledgments section is defined using the "acks" environment
%% (and NOT an unnumbered section). This ensures the proper
%% identification of the section in the article metadata, and the
%% consistent spelling of the heading.
\begin{acks}
This research is supported by the National Natural Science Foundation of China (9227010114, 62302321, 62372399, 62121002), the University Synergy Innovation Program of Anhui Province (GXXT-2022-040), and the advanced computing resources provided by the Supercomputing Center of Hangzhou City University.
\end{acks}

%%
%% The next two lines define the bibliography style to be used, and
%% the bibliography file.
\bibliographystyle{ACM-Reference-Format}
\bibliography{sample-base}

%%
%% If your work has an appendix, this is the place to put it.
% \appendix

\end{document}